\documentclass[aip,jcp,reprint,nofootinbib]{revtex4-1}
\usepackage{amsmath}       % Package  for subequations
\usepackage{graphicx}      % Package for figures
\usepackage{amssymb}       % For mathematical constructions
\usepackage{latexsym}      % Package to generate mathematical symbols
\usepackage{multirow}      % Self explnatory package for tables
\usepackage{array}         % Necessary to get the centering of objects to come out right in tables
\usepackage{natbib}		   % Allows inline citation options
\usepackage{hyperref}      % Package for hypertext links, external documents and URLs
\usepackage{placeins}
\newcommand{\bs}[1] {  \boldsymbol{#1}           }
\newcommand{\ff}[1] {  \mbox{\footnotesize{#1}}  }
\newcommand{\Ang}   {  \mbox{\normalfont\AA}     }
 
\begin{document}
%----------------------------------------------------------------------------------------
%	TITLE AND AUTHORS
%----------------------------------------------------------------------------------------
\title{The hydrogen bond network of water supports propagating optical phonon-like modes} % For titles, only capitalize the first letter
%% Enter authors via the \author command.  
%% Use \affil to define affiliations.
%% (Leave no spaces between author name and \affil command)

%% Note that the \thanks{} command has been disabled in favor of
%% a generic, reserved space for PNAS publication footnotes.

%% \author{<author name>
%% \affil{<number>}{<Institution>}} One number for each institution.
%% The same number should be used for authors that
%% are affiliated with the same institution, after the first time
%% only the number is needed, ie, \affil{number}{text}, \affil{number}{}
%% Then, before last author ...
%% \and
%% \author{<author name>
%% \affil{<number>}{}}

%% For example, assuming Garcia and Sonnery are both affiliated with
%% Universidad de Murcia:
%% \author{Roberta Graff\affil{1}{University of Cambridge, Cambridge,
%% United Kingdom},
%% Javier de Ruiz Garcia\affil{2}{Universidad de Murcia, Bioquimica y Biologia
%% Molecular, Murcia, Spain}, \and Franklin Sonnery\affil{2}{}}

%rev-tex version of authors: 
\author{Daniel C. Elton}
\email{daniel.elton@stonybrook.edu}
\affiliation{Department of Physics and Astronomy,  Stony Brook University, Stony Brook, New York 11794-3800, USA}
\affiliation{Institute for Advanced Computational Sciences, Stony Brook University, Stony Brook, New York 11794-3800, USA}

\author{Marivi Fern\'{a}ndez-Serra} 
\email{maria.fernandez-serra@stonybrook.edu}
\affiliation{Department of Physics and Astronomy,  Stony Brook University, Stony Brook, New York 11794-3800, USA}
\affiliation{Institute for Advanced Computational Sciences, Stony Brook University, Stony Brook, New York 11794-3800, USA}

%----------------------------------------------------------------------------------------
\begin{abstract}
The local structure of liquid water as a function of temperature is a source of intense research. This structure is intimately linked to the dynamics of water molecules, which can be measured using Raman and infrared spectroscopies. The assignment of spectral peaks depends on whether they are collective modes or single molecule motions. Vibrational modes in liquids are usually considered to be associated to the motions of single molecules or small clusters. Using molecular dynamics simulations we find dispersive optical phonon-like modes in the librational and OH stretching bands. We argue that on subpicosecond time scales these modes propagate through water's hydrogen bond network over distances of up to two nanometers. In the long wavelength limit these optical modes exhibit longitudinal-transverse splitting, indicating the presence of coherent long range dipole-dipole interactions, as in ice. Our results indicate the dynamics of liquid water have more similarities to ice than previously thought.
\end{abstract}
\maketitle

%----------------------------------------------------------------------------------------
%	PUBLICATION CONTENT
%----------------------------------------------------------------------------------------
The local structure and dynamics of liquid water as a function of temperature remains a source of intense research and lively debate.\cite{Santra2015:1,Errington2001:318,English2011:037801,Huang15214:2009,Mallamace2013:4899,Sahle2013:6301} 
A thus far unrecognised discrepancy exits between the peak assignments reported in Raman spectra with those reported in dielectric/IR spectra.
 Although early experimentalists fit the Raman librational band with two peaks,\cite{Walrafen1964:3249} it is better fit with three (Supplementary Table 1).\cite{Carey1998:2669,Walrafen1990:2237,Walrafen1967:114,Walrafen1986:6970,Castner1995:653} Previously these three peaks were assigned to the three librational motions of the water molecule - twisting ($\approx 435$ cm$^{-1}$), rocking ($\approx 600$ cm$^{-1}$) and wagging ($\approx 770$ cm$^{-1}$).\cite{Carey1998:2669,Walrafen1990:2237,Walrafen1986:6970} However, when comparing these assignments to infrared and dielectric spectra, one runs into a serious discrepancy. One expects to find the two higher frequency modes to be present, since only the rocking and wagging librations are IR active. The twisting libration, consisting of a rotation of the hydrogen atoms around the C2 axis, is not IR active since it does not affect the dipole moment of the molecule. Instead, IR spectra show two peaks at 380 and 665 cm$^{-1}$,\cite{Zelsmann1995:95} and similarly dielectric spectra show peaks at 420 and 620 cm$^{-1}$,\cite{Fukasawa2005:197802} in disagreement with this assignment.

The assignment of longitudinal optical phonon modes to Raman spectra can be made by looking at the longitudinal dielectric susceptibility. This method has been used previously to assign longitudinal phonon modes to the Raman spectra of ice Ih,\cite{Aure1978:65,Klug1991:7011,Whalley1977:3429} ice Ic,\cite{Klug1978:55} and vitreous GeO$_2$ and SiO$_2$.\cite{Galeener1976:1474}
It has previously been shown that the librational peak in the longitudinal dielectric susceptibility of water is dispersive,\cite{Restat:7277} and Bopp \& Kornyshev noted that the dispersion relation has the appearance of an optical phonon mode.\cite{Kornyshev1998:1939} The longitudinal mode in the dielectric susceptibility is equivalent to the dispersive mode discovered by Ricci et al. (1989) in the spectrum of hydrogen density fluctuations.\cite{Ricci1989:7226} 

Comparison of peak positions in longitudinal and transverse dielectric susceptibilities often reveals longitudinal-transverse (LO-TO) splitting. LO-TO splitting indicates the presence of long-range dipole-dipole interactions in the system. One way to understand LO-TO splitting is through the Lyddane-Sachs-Teller (LST) relation:\cite{Lyddane1941:673}
\begin{equation}
	\frac{\omega_{\ff{LO}}^2}{\omega_{\ff{TO}}^2} = \frac{\varepsilon(0)}{\varepsilon_\infty}
\end{equation}
Although this relation was originally derived for a cubic ionic crystal it was later shown to have very general applicability,\cite{Barker1975:4071,Sievers1990:3455} and has been applied to disordered and glassy solids.\cite{Whalley1977:3429,Payne1984:351,Sekimoto1982:3411} To apply this equation to water we must use a generalized LST  relation which takes into account all of the optically active modes in the system and the effects of dampening.\cite{Barker1975:4071} The generalized LST relation is:\cite{Barker1975:4071}
\begin{equation}\label{gLST}
	\prod_i \frac{\omega_{\ff{LDi}}}{\omega_{\ff{TDi}}} \prod_j \frac{|\bar{\omega}_{\ff{Lj}}|^2}{\omega_{\ff{Tj}}^2}= \frac{\varepsilon(0)}{\varepsilon_\infty}
\end{equation}
Here the index $i$ runs over the Debye peaks in the system and the index $j$ runs over the number of damped harmonic oscillator peaks. The longitudinal frequencies of the damped harmonic oscillators must be considered as complex numbers ($\bar{\omega}_{Li} = \omega_{Li} + i\gamma_i$), where $\gamma_i$ is the dampening factor. 

As shown by Barker, the generalized LST equation can be understood purely from a macroscopic point of view,\cite{Barker1975:4071} so by itself it yields little insight into microscopic dynamics. LO-TO splitting can be understood from a microscopic standpoint via the equation:\cite{Decius1977,Decius1968:1387}
\begin{equation}\label{LSTdipole}
	\omega_{Lk}^2 - \omega_{Tk}^2 = \frac{4\pi C}{3 v}  \left(\frac{\partial\bs{\mu}}{\partial{Q_k}}\right)^2 
\end{equation} 
Here $v$ is the volume per unit cell, $Q_k$ is the normal coordinate of mode $k$, and $C$ is a prefactor which depends on the type of lattice and the boundary conditions of the region being considered (for an infinite cubic lattice, $C$ = 1). %In particular, $C$ can be related to the Lorentz field factor, which has been tabulated for many crystals.\cite{Decius1968:1387}\cite{Mueller1935:947}
Equation \ref{LSTdipole} shows that LO-TO splitting is intimately related to crystal structure, and it has been used to 
evaluate the quasi-symmetry of room temperature ionic liquids.\cite{Burba2011:134503}.

In this work we show how the dielectric susceptibility can be used to probe water's local structure and dynamics. Our work solves the aforementioned peak assignment discrepancy. We find that the lowest frequency librational Raman peak ($\approx 435$ cm$^{-1}$) is a transverse optical phonon-like mode while the highest frequency peak ($\approx 770$ cm$^{-1}$) is a longitudinal optical phonon-like mode. This explains why the highest frequency Raman mode does not appear in IR or dielectric experiments, since such experiments only report the transverse response. We show that the transverse counterpart also exhibits dispersion. We argue that these dispersive modes are due to optical phonons that travel along the H-bond network of water. Our results indicate that not only does water exhibit LO-TO splitting, but also that its dependence with temperature is anomalous. We suggest that this measurement provides an alternative probe to evaluate structural changes in liquid water as a function of temperature.

%-------------------------------------------------------------------------------------------------------
\section{Results}
\begin{figure}
   \includegraphics[width=8.5cm]{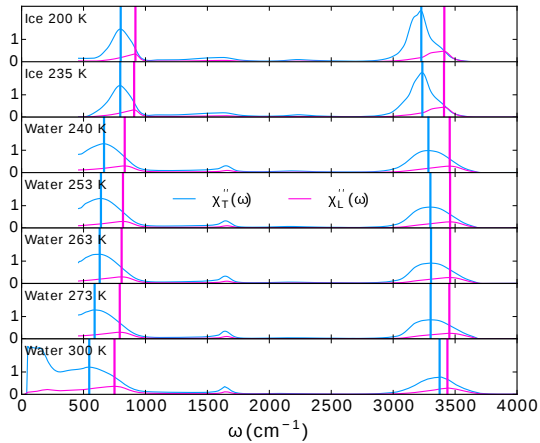}
    \caption{{\bf Dielectric susceptibilities of ice and water.} Computed from index of refraction data using equations \ref{epsL} and \ref{indexrelation}. data from 210 to 280 K comes from aerosol droplets\cite{Zasetsky2005:2760} while the data at 300 comes from bulk liquid.\cite{Hale:73}}
  	\label{chiLchiTexpt}
\end{figure}
\begin{figure}
  \includegraphics[width=8.5cm]{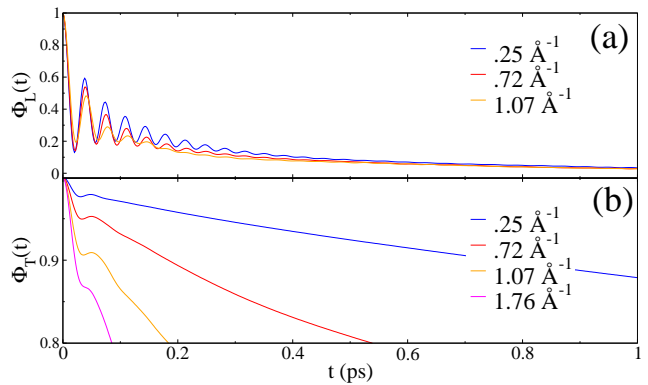}
  \caption{{\bf Polarization correlation functions.} Longitudinal (a) and transverse (b) polarization correlation functions (see equation \ref{PolCorrFns}) for TIP4P/$\varepsilon$, a rigid model. The oscillations at small $k$ come from the collective librational mode, which is much more pronounced in the longitudinal case.}\label{PhiCombined}
\end{figure}

As in our previous work\cite{Elton2014:124504} we compared results from a rigid (TIP4P/$\varepsilon$) model, a flexible model (TIP4P/2005f), and a flexible and polarizable model (TTM3F) in all of our analyses.

\section{LO-TO Splitting from experimental data}
We wish to study the $k$ dependence of the dielectric susceptibility, where $k=2\pi/\lambda$. $k-$ dependence cannot be probed directly by experiment, but in the limit of infinite wavelength ($k\rightarrow 0$) the longitudinal and transverse dielectric susceptibilities can be obtained from the dielectric function via the following relations:\cite{Madden2007:467,Hansen2006:341} 
\begin{equation}\label{epsL}
  \chi_L(k\rightarrow 0, \omega) =  1 - \frac{1}{\varepsilon(\omega)}
\end{equation}
\begin{equation}\label{epsT}
 \chi_T(k \rightarrow 0, \omega) =  \varepsilon(\omega)- 1 
\end{equation}
%\begin{equation}
%		\chi_L(k,\omega) = 1 - \frac{1}{\varepsilon_L(k,\omega) }
%\end{equation}
%\begin{equation}	    \chi_T(k,\omega) =  \varepsilon_T(k,\omega) - 1
%\end{equation}
Note that the transverse susceptibility is what one normally calls susceptibility. The dielectric function can be obtained from the index of refraction $n(\omega)$ and extinction coefficient $k(\omega)$ as: 
\begin{equation}\label{indexrelation}
	\begin{aligned}
		\varepsilon'(\omega)  &= n^2(\omega) - k^2(\omega)\\
		\varepsilon''(\omega) &= 2n(\omega)k(\omega) 
	\end{aligned}
\end{equation}
These equations allow us to use previously published experimental data\cite{Zasetsky2005:2760,Hale:73} to calculate the imaginary part of the longitudinal response. We find significant LO-TO splitting in the librational and stretching bands (fig. \ref{chiLchiTexpt}). 

%---------------------------------------------------------------
\subsection{Polarization correlation functions}
The normalized longitudinal and transverse polarization correlation functions are defined as: 
\begin{equation}\label{PolCorrFns}
	\Phi_{L/T}(k,t) \equiv \frac{ \langle \bs{P}_{L/T}(k,t)\cdot\bs{P}_{L/T}^*(k,0)\rangle }{  \langle \bs{P}_{L/T}(k,0)\cdot\bs{P}_{L/T}^*(k,0)\rangle }\\
\end{equation}
The correlation functions found for TIP4P/$\varepsilon$ at small small $k$ are shown in figure \ref{PhiCombined}. Since TIP4P/$\varepsilon$ is a rigid model, only librational motions are present. The addition of flexibility and polarizability add additional high frequency oscillations to the picture (Supplementary Fig. 1). In the small wavenumber regime ($k < 1.75 \Ang$) there is a damped oscillation which corresponds to the collective librational phonon-like mode. This damped oscillation is superimposed on an underlying exponential relaxation in both the transverse and longitudinal cases. In the longitudinal case the  relaxation time $\tau(k)$ of the underlying exponential relaxation exhibits non-monotonic behaviour with $k$, reaching a maximum at $k \approx 3 \Ang^{-1}$ (Supplementary Fig. 2). At wavenumbers greater than $k \approx 2.5 \Ang $ only intramolecular motions contribute.

%-------------------------------------------------------------------------------------------------------
\subsection{Dispersion of the librational peak}
\begin{figure}
  \includegraphics[width=7.5cm]{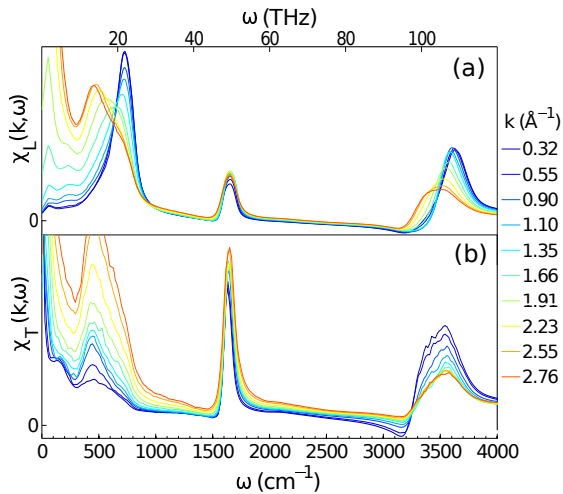}
  \caption{{\bf Imaginary part of longitudinal (a) \& transverse (top) susceptibility} From a simulation with TTM3F at 300 K. In the longitudinal spectra both the librational ($\approx$ 750 cm$^{-1}$) and OH stretching peak ($\approx$ 3500 cm$^{-1}$) peaks exhibit dispersion.}\label{lineplotTTM3F}
\end{figure}
\begin{table}\label{ObservedFreqs}
    \begin{tabular}{@{\vrule height 8pt depth 0pt width 0 pt} c c c c c c c c }    
Model & Temp & $\omega_{LO}$ &$\tau_{LO}$ & $\omega_{TO}$ & $\tau_{TO}$ &  $\omega_{LO} - \omega_{TO}$ \\ 
 \hline \multirow{4}{*}{TIP4P/2005f} 
    & 250  & 905  & .38   & 667  & .23   & 233 &    \\ %T=2139 m/s
    & 300	& 900  & .44   & 632  & .18  & 268 &    \\ 
    & 350	& 871  & .34   & 574  & .18  & 297 &    \\ 
 	& 400	& 826  & .25   & 423  & .17  & 400 &    \\ 
 \hline 
 \multirow{3}{*}{TTM3F }
    & 250	& 757  & .49   & 496  &      & 261 &   \\  
    & 300   & 721  & .44   & 410  &      & 311 &   \\ 
    & 350	& 710  & .20   & 380  &      & 330 &   \\ 
  \hline
expt\cite{Zasetsky2005:2760}& 253   & 820  &       & 641  &      & 179 &   \\ %1.19/1.77 
%expt\cite{Zasetsky2005:2760}& 253    & 690  &       & 502  &      & 188 &   \\ %1.19/1.77 %(fit Brendel) %expt\cite{Zasetsky2005:2760}& 253    & 702  &       & 520  &      & 188 &   \\ %1.19/1.77 %(fit DHO) 
expt\cite{Hale:73}  	    & 300    & 759  &       & 556  &      & 203 &   \\ %5.8/1.77 
%expt\cite{Hale:73}  	        & 300    & 723  &       & 558  &      & 176 &   \\ %5.8/1.77 %(fit DHO) 
      \end{tabular}
          \caption{{\bf Resonance frequencies and lifetimes} Frequencies are given in cm$^{-1}$ and lifetimes in picoseconds. The values from simulation were computed at the smallest $k$ in the system. The experimental values are based on the position of the max of the band and therefore only approximate.}
\end{table}
Figure \ref{lineplotTTM3F} shows the imaginary part of the longitudinal and transverse susceptibility for TTM3F. In the longitudinal case the librational peak is clearly seen to shift with $k$. In the transverse case, the lower frequency portion of the band is seen to shift slightly with $k$. Dispersion relations for the longitudinal and transverse librational peaks are shown in figure \ref{dispersionrelations} for three different temperatures, using one peak fits. The dispersion relations appear to be that of optical phonons. In both the longitudinal and transverse case the dampening factors remain less than the resonance frequencies, indicating an underdamped oscillation (Supplementary Fig. 3). The longitudinal dispersion relation for TIP4P/2005f agrees with that found by Bopp \& Kornyshev (who used the flexible BJH model).\cite{Kornyshev1998:1939} Resat et al. also obtained a similar dispersion relation (but at a higher frequency), using the reference memory function approximation for TIP4P instead of molecular dynamics.\cite{Restat1992:2618} %(855 cm$^{-1}$)

Resonance frequencies and lifetimes for the smallest $k$ are shown in table \ref{ObservedFreqs}. The speed of propagation of these modes was computed by finding the slope $d\omega/dk$ in the regime of linear dispersion. For TIP4P/2005f we found speeds of $\approx 2700$ m/s and $\approx 1800 $ m/s for the longitudinal and transverse modes. These propagation speeds are above the speed of sound in water (1500 m/s) but below the speed of sound in ice (4000 m/s). The temperature dependence of the propagation speed was found to be very small. 
 
In both the longitudinal and transverse cases, the residual of the peak fitting shows features not captured by our Debye + one resonance fit of the librational peak. In both the longitudinal and transverse cases there is a non-dispersive peak at higher frequency, located at $\approx $900 cm$^{-1}$ for TIP4P/2005f and at $\approx $650 cm$^{-1}$ in TTM3F. This peak is negligibly small in the $k=0$ longitudinal susceptibility but appears as a shoulder as $k$ increases. In the transverse case the overlapping peak persists at $k=0$, so we found that the $k=0$ transverse spectra is best fit with two peaks, in agreement with experimental spectra. As we describe later, the higher frequency transverse peak is largely due to the self part of the response and is associated with the wagging librations of single molecules. 

%-------------------------------------------------------------------------------------------------------
\subsection{Importance of polarizability}
\begin{figure}
\includegraphics[width=8.5cm]{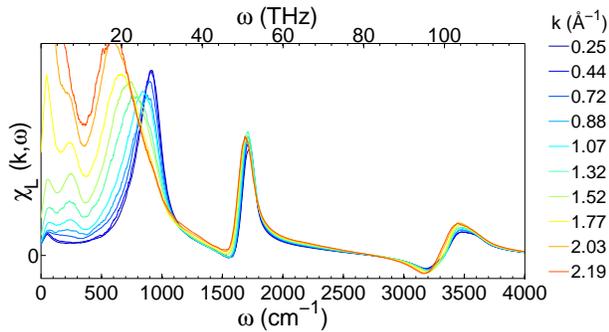}
  \caption{{\bf Imaginary part of the longitudinal susceptibility.} For TIP4P/2005f at 300 K. No significant dispersion is observed in the OH stretching peak. }\label{lineplot}
\end{figure}

There are several notable differences between TTM3F and the non-polarizable model TIP4P/2005f. First of all, the librational band of TIP4P/2005f is at higher frequency, in worse agreement with experiment. This difference in frequency is likely related to the parameters of TIP4P/2005f and not its lack of polarization. More importantly, we find that TTM3F exhibits dispersion in the OH stretching band ($\approx 3500$ cm$^{-1}$) in the longitudinal case while TIP4P/2005f does not. The transverse susceptibility of TTM3F does not exhibit such dispersion but the magnitude of the OH stretching band increases at small $k$, indicating long range intermolecular correlations. TIP4P/2005f does not exhibit this behavior. Similarly, at $k=0$ TTM3F exhibits significant LO-TO splitting in the OH stretching band while TIP4P/2005f does not (fig. \ref{chiLchiTvsexpt}). These findings are consistent with Heyden et al.'s results for the $k$-resolved IR spectra from ab-initio simulation, where they concluded that polarization allows for intermolecular correlations at the OH-stretch frequency.\cite{Heyden2012:2135} 
	
These findings can be understood from the dipole derivative in equation \ref{LSTdipole}. In the librational band the derivative of the dipole moment with respect to normal coordinate is purely due to rotation, while in the OH-stretching band it is due to changes in the geometry of the molecule and electronic polarization of the molecule during the OH stretching. In principle there may be coupling between the librational and stretching motions, but typically such rotational-vibrational coupling effects are negligibly small.\cite{woodward1972} The dipole moment surface (fluctuating charges) and polarization dipole incorporated in TTM3F account for the changes in polarization that occur during OH stretching motion. These results confirm the significance of polarization in capturing the OH stretching response of water.\cite{Heyden2012:2135} 

Figure \ref{chiLchiTvsexpt} shows a comparison of TTM3F, TIP4P/2005f and experiment at $k=0$. While the location of the peaks in TTM3F are in good agreement with the experimental data at 298 K, the magnitude of the longitudinal response is greatly overestimated in TTM3F. The degree of LO-TO splitting in the OH stretching peak is also overestimated in TTM3F. In general it appears that TTM3F overestimates the dipole derivative in equation \ref{LSTdipole} while TIP4P/2005f underestimates it. Figure \ref{chiLchiTvsexpt} also shows the effect of polarization at low frequencies, in particular the appearance of an H-bond stretching response at $\approx$ 250 cm$^{-1}$ in TTM3F which is absent in TIP4P/2005f.\cite{Elton2014:124504}

%-------------------------------------------------------------------------------------------------------
\subsection{LO-TO splitting vs temperature}
\begin{figure}
 \includegraphics[width=8.5cm]{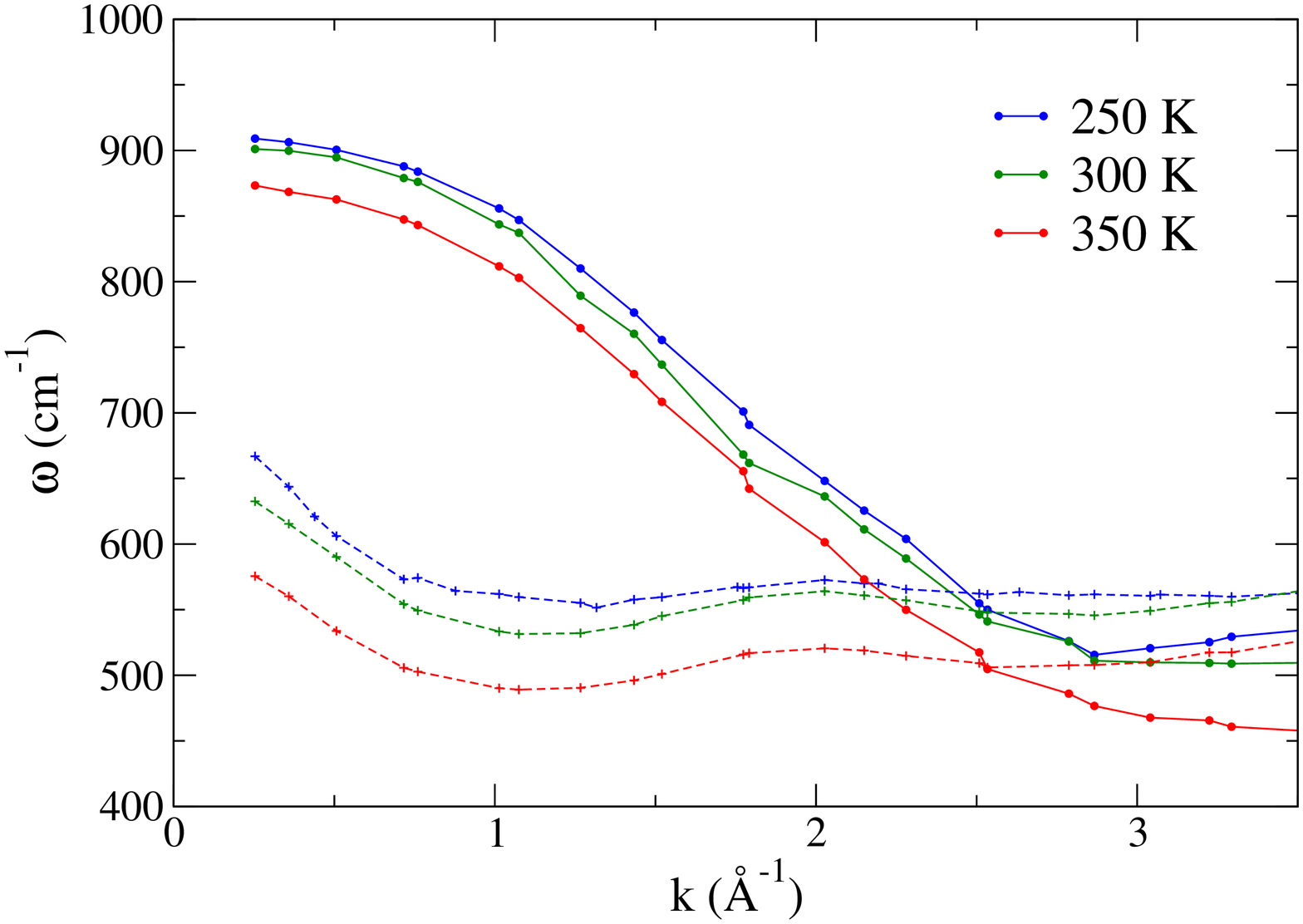}
  \caption{{\bf Dispersion relations for the propagating librational modes.} For TIP4P/2005f at three different temperatures (squares = longutudinal, pluses = transverse). A similar plot was found for TTM3F, but with lower frequencies.}\label{dispersionrelations}
\end{figure}

The frequencies of the librational and stretching modes are shown in table \ref{ObservedFreqs}. Once again we compare our results to experimental data.\cite{Wagner2005:7099,Hale:73,Zasetsky2005:2760} The comparison is imperfect since the TIP4P/2005f and TTM3F data comes from data at finite $k$ (the smallest $k$ in the system). For all three systems (TIP4P/2005f, TTM3F, and experiment) the increase in the LO-TO splitting of the librational band is puzzling, since the right hand side of the LST relation predicts a decrease in splitting, corresponding to a smaller dielectric constant and weaker dipole-dipole interactions. We found verifying the generalized LST equation is difficult because water contains either two or three Debye relaxations which must be taken into account.\cite{Vinh2015:164502,Ellison2007:1} Uncertainties in how to fit the region of 1 - 300 cm$^{-1}$ (.2- 9 THz), which includes contributions from many H-bonding modes, precludes a direct application of the generalized LST relation to water. By ignoring this region, however, we were able to achieve an approximate validation of the generalized LST equation for TIP4P/2005f. A more detailed analysis of how to fit the low frequency region will be the focus of future work. Since the generalized LST equation is an exact sum rule it can be used to assist in testing the validity of different fit functions. 

%-------------------------------------------------------------------------------------------------------
\subsection{Relation to phonons in ice}
\begin{figure}
  \includegraphics[width=8.5cm]{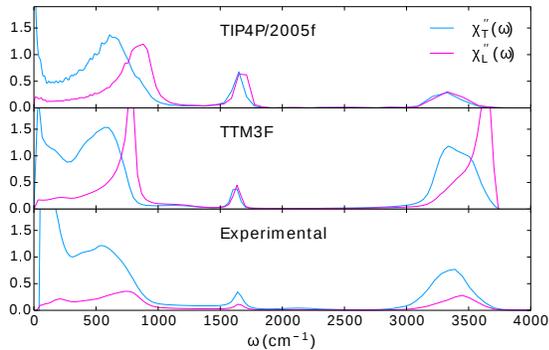}
  \caption{{\bf Imaginary parts of dielectric susceptibility.} We compare (a) the non-polarizable model TIP4P/2005f, (b) the polarizable model TTM3F, and (c) experimental data\cite{Hale:73} at 298 K. The effects of polarization can be seen in the LO-TO splitting of the stretching mode and in the low frequency features.}
  \label{chiLchiTvsexpt}
\end{figure}
Naturally we would like to find corresponding optical phonon modes in ice. As shown in figure \ref{chiLchiTexpt} the dielectric spectra and LO-TO splitting of supercooled water resembles that of ice. Recently evidence has been presented for propagating librational phonon modes in ice XI.\cite{Iwano2010:063601,Shigenari2012:174504} Three of the twelve librational modes of ice XI are IR active (labeled WR1, RW1 and RW2) and all three exhibit LO-TO splitting. The splittings have been found from Raman scattering to be 255, 135 and 35 cm$^{-1}$.\cite{Shigenari2012:174504} These modes all consist of coupled wagging and rocking motions. The WR1 mode, which has the largest infrared intensity, most closely matches our results. WR1 and RW2 have the same transverse frequency and RW1 has a smaller infrared intensity, which may help explain why the librational band is well fit by a single optical mode. LO-TO splitting in the OH-stretching modes of ice Ih has been discussed previously.\cite{Whalley1977:3429}
 
%-------------------------------------------------------------------------------------------------------
\subsection{Range of propagation}
\begin{figure}
  \includegraphics[width=8.5cm]{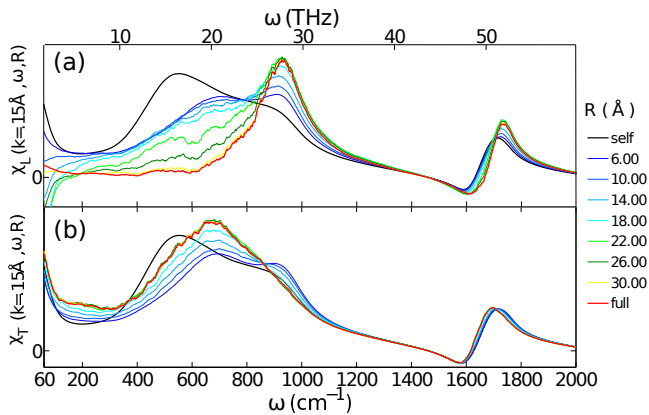}
  \caption{{\bf Imaginary part of the distance decomposed susceptibility for TIP4P/2005f.} Transverse (a) and longitudinal (b) susceptibilities, calculated with a 4 nm box at 300K, using the smallest $k$ vector in the system. Gaussian smoothing was applied. Long range contributions to the librational peak extending to $R = 2$ nm are observed.}\label{DistDep}
\end{figure}

The range of propagation of these modes can be calculated as $R = \tau v_g$ where $\tau$ is the lifetime and $v_g = d\omega/dk$ is the group velocity. For TIP4P/2005f we find a range of propagation of $\approx$ 1.1 nm for the longitudinal librational mode and $\approx$ .3 nm for the transverse mode. Similar results hold for TTM3F.

To verify that the modes we observe are actually propagating and to further quantify the range of propagation we study the spatial extent of polarization dipole correlations as a function of frequency. We investigated several different methodologies that can be used to decompose a spectra into distance-dependent components (Supplementary Note 1). We choose to start with the polarization correlation function: 
\begin{equation}
	\phi(k,t) = \left\langle \sum_i \bs{p}_i(k,0) \cdot \sum_j \bs{p}_j(k,t) \right\rangle
\end{equation} 
Here $\bs{p}_i(k,t)$ represents either the longitudinal or transverse molecular polarization vector of molecule $i$. We now limit the molecules in the second sum to those in a sphere of radius $R$ around each molecule $i$:
\begin{equation}
\phi(k,t,R) = \left\langle \sum_i \bs{p}_i(k,0) \cdot \sum_{j \in R_i} \bs{p}_j(k,t) \right\rangle
\end{equation}
The resulting function exhibits the expected $R \rightarrow 0$ limit, yielding only the self contribution. $R$ can be increased to the largest $R$ in the system ($\sqrt{3}L/2$), where the full response function for the simulation box is recovered. As $R$ increases, the contributions of the distinct term add constructively and destructively to the self term, illustrating the contributions from molecules at different distances. 
 
Figure \ref{DistDep} shows the distance decomposed longitudinal and transverse susceptibilities for TIP4P/2005f in a 4 nm box at the smallest $k$ available in the system. The entire region between 0 - 1000 cm$^{-1}$ contains significant cancellation between the self and distinct parts, in qualitative agreement with a previous study.\cite{Wan2013:4124} In the longitudinal susceptibility, the self component has two peaks (at 500 and 900 cm$^{-1}$ for TIP4P/2005f) representing the two IR active librational motions (rocking and wagging, respectively). The self part is the same in both the longitudinal and transverse cases, reflecting an underlying isotropy which is only broken when dipole-dipole correlations are introduced. 
%One can also note that the $R$ dependent results match nicely with the $k$ dependent results - the largest $R$ correspond to the smallest $k$. 
Further insight into the self-distinct cancellation comes from the results of Bopp, et. al., who project the hydrogen currents into a local molecular frame, allowing them to study the cross correlations between the rocking and wagging librations.\cite{Kornyshev1998:1939} They find that  in the longitudinal case cross correlations between rocking and wagging contribute negatively in the region of 480 cm$^{-1}$ and positively in the region of 740 cm$^{-1}$, suppressing the lower frequency peak to zero and enhancing the higher frequency peak.

In both the transverse and longitudinal cases as $R$ increases a new peak emerges, corresponding to the propagating mode. Incidentally, the shift in the peak between the self and distinct parts rules out the possibility that the propagating mode is the proposed dipolar plasmon resonance since the dipolar plasmon must be a resonance of both the of single molecule and collective motion.\cite{Lobo1973:5992,Pollock1981:950,Chandra1990:6833} Interestingly, there are very long range contributions to this peak. In our simulations with a 4 nm box of TIP4P/2005f contributions persist up to 3 nm in the longitudinal case and 2 nm in the transverse case. As noted, recent studies of ice XI suggest that the propagating modes consist of coupled wagging and rocking librations.\cite{Iwano2010:063601,Shigenari2012:174504} The results for the transverse mode seem to confirm this hypothesis for liquid water, since the propagating mode peak lies between the single molecule rocking and wagging peaks. In the longitudinal case the propagating mode overlaps more with the wagging peak, suggesting a greater role for these type of librations in the longitudinal phonon. 

%The distance dependence decomposition shows the actual situation is more complex - the lower frequency peak corresponds to the propagating optical mode which is likely a mixture of both librations. The higher frequency peak on the other hand does correspond largely to single molecule (self part) wagging librations although it is modified somewhat by the distinct part. 

%-------------------------------------------------------------------------------------------------------
\subsection{Methanol \& acentonitrile}
\begin{figure}
 \includegraphics[width=8.5cm]{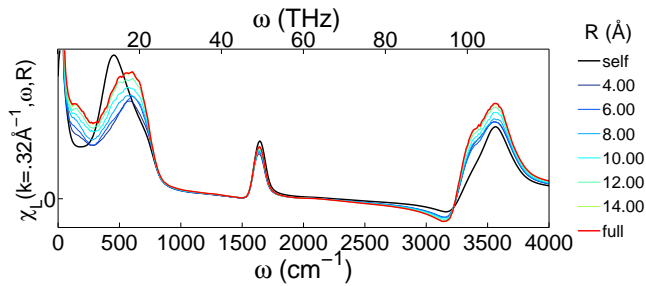}
  \caption{{\bf Imaginary part of the distance decomposed longitudinal susceptibility for TTM3F at 300 K.} Long range contributions are observed in the OH stretching band.}\label{DistDepTTM3F}
\end{figure}
To provide further evidence the aforementioned optical modes propagate through the hydrogen bond network of water we decided to repeat our analysis for other polar liquids, both H-bonding and non H-bonding. %Previously it has been shown that the longitudinal polarization correlation function of acetonitrile (a non H-bonding polar liquid) does not exhibit pronounced oscillations.\cite{Branka1999:250}
%Previous work using molecular dynamics simulation has shown that the time dependent solvation correlation functions of water and methanol (CH$_3$OH) exhibit fast oscillations, whereas acetonitrile (CH$_3$CN) and methyl chloride (CH$_3$Cl) do not.\cite{Maroncelli1994:310} This is not surprising since methanol contains an OH group which can undergo fast librational motion, while acetonitrile and methyl chloride do not. 
As an H-bonding liquid we choose methanol, which is known to contain winding H-bonded chains. According to results from MD simulation, most of these chains have around 5-6 molecules\cite{Haughhey1987:4934,Yamaguchi2000:8976}, with a small percentage of chains containing 10-20 molecules.\cite{Matsumoto1990:1981} Chain lifetimes have been estimated to be about .5 ps.\cite{Matsumoto1990:1981} Therefore we expect methanol can also support a librational phonon mode that propagates along hydrogen bonds, but perhaps with a shorter lifetime and range than water. As a non H-bonding polar liquid we choose acetonitrile, because it has a structure similar to methanol, but with the hydroxyl group replaced by a carbon atom. We find that the OH librational band of methanol ($\approx 700$ cm$^{-1}$\cite{Crowder1967:914}) is indeed dispersive (Supplementary Fig. 4). As with water, the transverse spectra also exhibits dispersion, but to a much lesser extent. LO-TO splitting of about 100 cm$^{-1}$ is observed in the 700 cm$^{-1}$ librational peak. The results for acetonitrile are more ambiguous - we observe dispersion in the broad peak at $\approx 100$ cm$^{-1}$, however this peak contains contributions from translational and (free) rotational modes, as well as the CH$_3$ torsion mode, and it is not clear which modes are responsible for the dispersion (Supplementary Fig. 5). 
%Experimental infrared spectra of ethanol - acetonitrile mixtures show a shifting of the methanol librational peak with increasing acetonitrile concentration.\cite{Venables2000:3243}  

%-------------------------------------------------------------------------------------------------------
\section{Discussion}
In this work we have presented several lines of evidence for short lived optical phonons that propagate along the H-bond network of water. The longitudinal and transverse nonlocal susceptibility exhibit dispersive peaks with dispersion relations resembling optical phonons. As the temperature is lowered, the resonance frequencies and LO-TO splittings of these modes converge towards the values for phonons in ice Ih. By comparing our results with a recent study of ice XI we believe both modes likely consist of coupled wagging and rocking librations.\cite{Iwano2010:063601}\cite{Shigenari2012:174504} 

This work fundamentally changes our understanding of the librational band in the Raman spectra of water by assigning the lower and higher frequency peaks to transverse and longitudinal optical modes. Our analysis of the self-distinct cancellation indicates that the middle Raman peak ($\approx 600$ cm$^{-1}$) belongs to the remnant of the single molecule wagging response which remains after the cancellation. We are also led to a new interpretation the librational region of the real part of the dielectric function. In the case of a lossless optical phonon the transverse phonon occurs where $\varepsilon'(\omega) = \infty$ while the longitudinal phonon occurs where $\varepsilon'(\omega) = 0$. The presence of dampening smooths the divergence leading to a peak followed by a sharp dip. This is what is observed in the real part of the dielectric function of water between 300 - 500 cm$^{-1}$ (the features are shifted to lower frequencies by the tail of the low frequency Debye relaxation). 

One might wonder how our work relates to existing work on acoustic modes in water, in particular, the controversial ``fast sound" mode.\cite{Santucci2006:225701,Sampoli1997:1678} Acoustic modes, which are observable through the dynamic structure factor, have been explored as means of understanding the hydrogen bond structure and low temperature anomalies of water.\cite{Mallamace2013:4899} In this work we have argued that optical modes can also provide insight into water's structure and dynamics. The fast sound mode lies at much lower frequencies than the librational and OH stretch modes that we studied. The H-bond bending and stretching modes also primarily lie at at frequencies below the librational region. However, normal mode analysis of liquid water and clusters shows that the H-bond stretching modes have a wide distribution of frequencies which overlaps with the librational modes, so some coupling between these modes is possible.\cite{Cho1994:6672,Garberoglio2002:3278} Recently it was shown that there is coupling between the acoustic and optic modes in water - ie. between fluctuations in mass density and fluctuations in charge density.\cite{Seldmeier2014:054512} 

The large spatial range and coherent propagation of these modes is surprising and implies the existence of an extended hydrogen bond network, in contrast to earlier ideas about the structure of water which emphasize dynamics as being confined within small clusters.\cite{Bosma1993:4413} Simulations with larger simulation boxes are needed to fully quantify the extent of the longitudinal modes. The ability of water to transmit phonon modes may be relevant to biophysics, where such modes could lead to dynamical coupling between biomolecules, a phenomena which is currently only being considered at much lower frequencies.\cite{Nibali2014:12800,Ebbinghaus2007:20749,Kim2008:6486} The methodology used in this paper to analyse LO-TO splitting opens up a new avenue to understanding the structure and dynamics of water. The fact that the librational LO-TO splitting increases with temperature instead of the expected decrease is likely due to significant changes in the structure of the liquid. One likely possibility is that the volume per ``unit cell" term in equation \ref{LSTdipole} decreases with temperature. This could be caused by the local quasi-structure determined by H-bonding changing from a more ice-like structure (4 molecules per unit cell) to a more cubic structure (1 molecule per unit cell). More research is needed to understand the microscopic origin of the LO-TO splitting in water, both in the librational and stretching modes.  

%-------------------------------------------------------------------------------------------------
\section{Methods}
\subsection{Theory of the nonlocal susceptibility}
If the external field is sufficiently small, the relation between the polarization response of a medium and the electric displacement field $\bs{D}$ for a spatially homogeneous system is given by :
\begin{equation}\label{generalexpr}
    \bs{P}(\bs{r},t) = \epsilon_0 \int_V \int_{-\infty}^{t} d\bs{r}' dt'  \stackrel{\leftrightarrow}{\bs{\chi}} (\bs{r} - \bs{r}', t - t') \bs{D}(\bs{r}',t')
\end{equation} 
%If the system is spatially homogeneous then ${\bs{\chi}} (\bs{r},\bs{r}', t - t') = {\bs{\chi}} (\bs{r} - \bs{r}', t - t')$ and the 
This expression Fourier transforms to: 
\begin{equation}
    \bs{P}(\bs{k},\omega) = \epsilon_0 \stackrel{\leftrightarrow}{\bs{\chi}}(\bs{k},\omega) \bs{D}(\bs{k},\omega) 
\end{equation}  
For isotropic systems, the tensor $\stackrel{\leftrightarrow}{\bs{\chi}}$ can be decomposed into longitudinal and transverse components: 
\begin{equation}
  \stackrel{\leftrightarrow}{\bs{\chi}}(\bs{k},\omega) =  \chi_L(k, \omega)\hat{\bs{k}}\hat{\bs{k}} + \chi_T(k, \omega)(\bs{I} - \hat{\bs{k}}\hat{\bs{k}}) 
\end{equation}

The easiest starting point for deriving microscopic expressions for $\chi_L(k,\omega)$ and $\chi_T(k, \omega)$ is the classical Kubo formula:\cite{Kubo1957:570} 
\begin{equation}\label{linearRT}
	\chi_{L/T}(\bs{k},\omega) = \frac{\beta}{\epsilon_0} \int_0^\infty dt \frac{d}{dt} \langle
\bs{P}_{L/T}(\bs{k},t)\cdot\bs{P}_{L/T}^*(\bs{k},0)\rangle e^{i\omega t}  
\end{equation}

This expression relates the susceptibility to the time correlation function of the polarization in equilibrium. The longitudinal part of the polarization can be calculated by Fourier transforming the defining expression for the polarization $\nabla\cdot\bs{P}(\bs{r},t)  = -\rho(\bs{r},t)$, leading to $\hat{\bs{k}}\cdot\bs{P} = i \rho(\bs{k},t)/k = P_L$. To calculate the transverse part of the polarization we use the method of Raineri \& Friedman to find the polarization vector of each molecule (Supplementary Note 2).\cite{Raineri1993:8910} 
%\begin{equation}
%	\Phi_L(k,t) \equiv \frac{ \langle \rho(\bs{k},t)\rho^*(\bs{k},0)\rangle }{  \langle %\rho(\bs{k},0)\rho^*(\bs{k},0)\rangle }\\
%\end{equation}
We can rewrite eqn. \ref{linearRT} in terms of the normalized polarization correlation function (eqn. \ref{PolCorrFns}), and taking into account the isotropy of water: 
\begin{equation}\label{correlation_fun_eqn}
	\chi_{L/T}(k,\omega) = \chi_{L/T}(k,0) \int_0^\infty \dot{\Phi}_{L/T}(k,t) e^{i\omega t} dt   
\end{equation}

\subsection{Computational methods}
The three water models we used were TIP4P/$\varepsilon$,\cite{Fuentes-Azcatl2014:1263} TIP4P/2005f,\cite{gonzalez:224516} and TTM3F.\cite{fanourgakis:074506} To simulate methanol and acetonitrile we used the General AMBER Forcefield (GAFF),\cite{Wang2004:1157} a forcefield with full intramolecular flexibility which has been shown to satisfactory reproduce key properties of both liquids.\cite{Caleman2012:61} Our TTM3F simulations were performed with an in-house code that uses the TTM3F force calculation routine of Fanourgakis and Xantheas. All other simulations were ran using the GROMACS package (ver. 4.6.5).\cite{Hess:435} We used particle-mesh Ewald summation for the long range electrostatics with a Coloumb cutoff of 2 nm for our 4+ nm simulations and a cutoff of 1.2 nm for our simulations with 512 molecules. Our TTM3F simulations had 256 molecules and used Ewald summation with a Coulomb cutoff of .9 nm. The principle TIP4P/2005f simulations contained 512 molecules and were 8 ns long ($\Delta t_{\ff{out}} = 8$ fs) and .6 - 1.2 ns long ($\Delta t_{\ff{out}} = 4$ fs). Other simulations were 1-2 ns long. Simulations with MeOH and ACN contained 1,000 molecules and were 1 ns long. All simulations were equilibrated for at least 50 ps prior to outputting data. 

Because of periodic boundary conditions, the possible $k$ vectors are limited to the form $\bs{k} = 2\pi n_x \hat{\bs{i}}/L_x + 2\pi n_y\hat{\bs{j}}/L_y + 2\pi n_z \hat{\bs{k}}/L_z$, where $n_x$, $n_y$, and $n_z$ are integers. We calculated correlation functions separately for each $\bs{k}$ and then average over the results for $\bs{k}$ vectors with the same magnitude, a process we found reduced random noise.

One can question whether a purely classical treatment is justified here because the librational dynamics we are interested have frequencies of 700-900 cm$^{-1}$ for which $\hbar\omega \approx 3-4 k_B T$ at 300 K. Previously it was shown that the widely-used harmonic correction does not change the spectrum.\cite{Kornyshev1998:1939} 
%whereas applying the so-called standard approximation does In either case, such quantum correction factors do not shift the position of peaks, but only change their magnitude (and to a lesser extent their shape).\cite{ramirez2004:39793} 
Furthermore, comparison of $k$ resolved IR spectra taken from molecular dynamics and ab-inito DFT simulation show that they give qualitatively similar results for all frequencies below 800 cm$^{-1}$.\cite{Heyden2012:2135} For the OH stretching peak, however, quantum effects are known to be very important.

%-----------------------------------------------------------------------------------
\subsection{Fitting the librational band}
To obtain resonance frequencies and lifetimes for the librational peak in the imaginary part of the response we used a damped oscillator model. A Debye peak overlaps significantly with the librational band in both the longitudinal and transverse cases and must be included in the peak fitting. Equation \ref{correlation_fun_eqn} can be used to relate the form of the time correlation function to the absorption peak lineshape. For Debye response one has the following expressions: 
\begin{equation}\label{nonresonant}
	\begin{aligned}
			    			\Phi(k,t) &= A e^{-t/\tau_D} \\
%\frac{\chi(k,\omega)}{\chi(k,0)} &= \frac{A}{1 - i\omega\tau_D}  \\ 
\frac{\mbox{Im}\lbrace \chi(k,\omega)\rbrace}{\chi(k,0)} &= \frac{A\omega\tau_D}{1  + \omega^2\tau_D^2} 
	\end{aligned}
\end{equation}
For resonant response with resonance frequency $\omega_0(k)$ and dampening factor $\gamma \equiv 1/\tau$ we have:
\begin{equation*}\label{resonant}
	\begin{aligned}
	    \Phi(k,t)      &= B e^{-t/\tau} \cos(\omega_0 t)   \\
% \frac{\chi(k,\omega)}{\chi(k,0)} &= \frac{B}{2}\left(\frac{1 - i\omega_0\tau}{1 - i(\omega + \omega_0)^2\tau}   + \frac{1 + i\omega_0\tau}{1 - i(\omega - \omega_0)^2\tau}\right) \\ 
\frac{\mbox{Im}\lbrace \chi(k,\omega)\rbrace}{\chi(k,0)} &= \frac{B}{2}\left(\frac{\omega\tau}{1 + (\omega + \omega_0)^2\tau^2} + \frac{\omega\tau}{1 + (\omega - \omega_0)^2\tau^2} \right)
	\end{aligned}
\end{equation*}
%This lineshape was first introduced by Van Vleck and Weisskopf.\cite{VanVleck1945:227}\cite{toda1991} 
%The Debye equations are seen to correspond to the case where $\omega_0 = 0$. 
%In contrast to the standard damped harmonic oscillator Lorentzian, the Van Vleck-Weisskopf lineshape is asymmetric with a maximum at $\omega_{\ff{max}} = \sqrt{1 + (\omega_0\tau)^2}/\tau$. 
We find this lineshape (the Van Vleck-Weisskopf lineshape\cite{toda1991,VanVleck1945:227}) yields results identical to the standard damped harmonic oscillator response for the range of $\tau, \omega_0$ values we are interested in. 
We found a two function (Debye + resonant) fit worked very well for fitting the librational peak in the longitudinal case (Supplementary Figs. 7 and 8). The H-bond stretching peak at $\approx$200 cm$^{-1}$ overlaps with the librational band for $2 < k < 2.5$, and we found that it can be included in the fit using an additional damped harmonic oscillator, but usually this was not necessary. Because of this overlap and due to the broad nature of the transverse band, the fitting in the transverse case is only approximate. We found this was especially true for TTM3F and the experimental data, so we do not report lifetimes for such cases.

%-------------------------------------------------------------------------------------------------------
\section{Acknowledgments}
This work was partially supported by DOE Award No. DE-FG02-09ER16052 (D.C.E.) and by DOE Early Career Award No. DE-SC0003871 (M.V.F.S.).

%----------------------------------------------------------------------------------------
%	BIBLIOGRAPHY
%----------------------------------------------------------------------------------------
\section*{References}
%\bibliography{../optical_phonons_paper.bib,dielectric.bib}

\begin{thebibliography}{10}
\expandafter\ifx\csname url\endcsname\relax
  \def\url#1{\texttt{#1}}\fi
\expandafter\ifx\csname urlprefix\endcsname\relax\def\urlprefix{URL }\fi
\providecommand{\bibinfo}[2]{#2}
\providecommand{\eprint}[2][]{\url{#2}}

\bibitem{Santra2015:1}
\bibinfo{author}{Santra, B.}, \bibinfo{author}{Jr., R. A.~D.},
  \bibinfo{author}{Martelli, F.} \& \bibinfo{author}{Car, R.}
\newblock \bibinfo{title}{Local structure analysis in ab initio liquid water}.
\newblock \emph{\bibinfo{journal}{Mol. Phys.}} \bibinfo{pages}{1--13}
  (\bibinfo{year}{in press 2015}).

\bibitem{Errington2001:318}
\bibinfo{author}{Errington, J.~R.} \& \bibinfo{author}{Debenedetti, P.~G.}
\newblock \bibinfo{title}{Relationship between structural order and the
  anomalies of liquid water}.
\newblock \emph{\bibinfo{journal}{Nature}} \textbf{\bibinfo{volume}{409}},
  \bibinfo{pages}{318--321} (\bibinfo{year}{2001}).

\bibitem{English2011:037801}
\bibinfo{author}{English, N.~J.} \& \bibinfo{author}{Tse, J.~S.}
\newblock \bibinfo{title}{Density fluctuations in liquid water}.
\newblock \emph{\bibinfo{journal}{Phys. Rev. Lett.}}
  \textbf{\bibinfo{volume}{106}}, \bibinfo{pages}{037801}
  (\bibinfo{year}{2011}).

\bibitem{Huang15214:2009}
\bibinfo{author}{Huang, C.} \emph{et~al.}
\newblock \bibinfo{title}{The inhomogeneous structure of water at ambient
  conditions} \textbf{\bibinfo{volume}{106}}, \bibinfo{pages}{15214--15218}
  (\bibinfo{year}{2009}).

\bibitem{Mallamace2013:4899}
\bibinfo{author}{Mallamace, F.}, \bibinfo{author}{Corsaro, C.} \&
  \bibinfo{author}{Stanley, H.~E.}
\newblock \bibinfo{title}{Possible relation of water structural relaxation to
  water anomalies}.
\newblock \emph{\bibinfo{journal}{Proc. Natl. Acad. Sci. USA}}
  \textbf{\bibinfo{volume}{110}}, \bibinfo{pages}{4899--4904}
  (\bibinfo{year}{2013}).

\bibitem{Sahle2013:6301}
\bibinfo{author}{Sahle, C.~J.} \emph{et~al.}
\newblock \bibinfo{title}{Microscopic structure of water at elevated pressures
  and temperatures}.
\newblock \emph{\bibinfo{journal}{Proc. Natl. Acad. Sci. USA}}
  \textbf{\bibinfo{volume}{110}}, \bibinfo{pages}{6301--6306}
  (\bibinfo{year}{2013}).

\bibitem{Walrafen1964:3249}
\bibinfo{author}{Walrafen, G.~E.}
\newblock \bibinfo{title}{Raman spectral studies of water structure}.
\newblock \emph{\bibinfo{journal}{J. Phys. Chem.}}
  \textbf{\bibinfo{volume}{40}}, \bibinfo{pages}{3249} (\bibinfo{year}{1964}).

\bibitem{Carey1998:2669}
\bibinfo{author}{Carey, D.~M.} \& \bibinfo{author}{Korenowski, G.~M.}
\newblock \bibinfo{title}{Measurement of the {R}aman spectrum of liquid water}.
\newblock \emph{\bibinfo{journal}{J. Chem. Phys.}}
  \textbf{\bibinfo{volume}{108}}, \bibinfo{pages}{2669--2675}
  (\bibinfo{year}{1998}).

\bibitem{Walrafen1990:2237}
\bibinfo{author}{Walrafen, G.~E.}
\newblock \bibinfo{title}{Raman spectrum of water: transverse and longitudinal
  acoustic modes below $\approx$ 300 cm$^{-1}$ and optic modes above $\approx$
  300 cm$^{-1}$}.
\newblock \emph{\bibinfo{journal}{J. Phys. Chem.}}
  \textbf{\bibinfo{volume}{94}}, \bibinfo{pages}{2237--2239}
  (\bibinfo{year}{1990}).

\bibitem{Walrafen1967:114}
\bibinfo{author}{Walrafen, G.~E.}
\newblock \bibinfo{title}{Raman spectral studies of the effects of temperature
  on water structure}.
\newblock \emph{\bibinfo{journal}{J. Phys. Chem.}}
  \textbf{\bibinfo{volume}{47}}, \bibinfo{pages}{114--126}
  (\bibinfo{year}{1967}).

\bibitem{Walrafen1986:6970}
\bibinfo{author}{Walrafen, G.~E.}, \bibinfo{author}{Fisher, M.~R.},
  \bibinfo{author}{Hokmabadi, M.~S.} \& \bibinfo{author}{Yang, W.}
\newblock \bibinfo{title}{Temperature dependence of the low‐ and
  high‐frequency {R}aman scattering from liquid water}.
\newblock \emph{\bibinfo{journal}{J. Chem. Phys.}}
  \textbf{\bibinfo{volume}{85}}, \bibinfo{pages}{6970--6982}
  (\bibinfo{year}{1986}).

\bibitem{Castner1995:653}
\bibinfo{author}{Castner, E.~W.}, \bibinfo{author}{Chang, Y.~J.},
  \bibinfo{author}{Chu, Y.~C.} \& \bibinfo{author}{Walrafen, G.~E.}
\newblock \bibinfo{title}{The intermolecular dynamics of liquid water}.
\newblock \emph{\bibinfo{journal}{J. Chem. Phys.}}
  \textbf{\bibinfo{volume}{102}}, \bibinfo{pages}{653--659}
  (\bibinfo{year}{1995}).

\bibitem{Zelsmann1995:95}
\bibinfo{author}{Zelsmann, H.~R.}
\newblock \bibinfo{title}{Temperature dependence of the optical constants for
  liquid {H}$_2${O} and {D}$_2${O} in the far {IR} region}.
\newblock \emph{\bibinfo{journal}{J. Mol. Str.}}
  \textbf{\bibinfo{volume}{350}}, \bibinfo{pages}{95--114}
  (\bibinfo{year}{1995}).

\bibitem{Fukasawa2005:197802}
\bibinfo{author}{Fukasawa, T.} \emph{et~al.}
\newblock \bibinfo{title}{Relation between dielectric and low-frequency {R}aman
  spectra of hydrogen-bond liquids}.
\newblock \emph{\bibinfo{journal}{Phys. Rev. Lett.}}
  \textbf{\bibinfo{volume}{95}}, \bibinfo{pages}{197802}
  (\bibinfo{year}{2005}).

\bibitem{Aure1978:65}
\bibinfo{author}{Aure, P.} \& \bibinfo{author}{Chosson, A.}
\newblock \bibinfo{title}{The translational lattice-vibration {R}aman spectrum
  of single crystal ice 1h}.
\newblock \emph{\bibinfo{journal}{J. Glaciology}}
  \textbf{\bibinfo{volume}{21}}, \bibinfo{pages}{65--71}
  (\bibinfo{year}{1978}).

\bibitem{Klug1991:7011}
\bibinfo{author}{Klug, D.~D.}, \bibinfo{author}{Tse, J.~S.} \&
  \bibinfo{author}{Whalley, E.}
\newblock \bibinfo{title}{The longitudinal‐optic–tranverse‐optic mode
  splitting in ice {I}h}.
\newblock \emph{\bibinfo{journal}{J. Chem. Phys.}}
  \textbf{\bibinfo{volume}{95}}, \bibinfo{pages}{7011--7012}
  (\bibinfo{year}{1991}).

\bibitem{Whalley1977:3429}
\bibinfo{author}{Whalley, E.}
\newblock \bibinfo{title}{A detailed assignment of the o–h stretching bands
  of ice i}.
\newblock \emph{\bibinfo{journal}{Canadian Journal of Chemistry}}
  \textbf{\bibinfo{volume}{55}}, \bibinfo{pages}{3429--3441}
  (\bibinfo{year}{1977}).

\bibitem{Klug1978:55}
\bibinfo{author}{Klug, D.~D.} \& \bibinfo{author}{Whalley, E.}
\newblock \bibinfo{title}{Origin of the high-frequency translational bands of
  ice i*}.
\newblock \emph{\bibinfo{journal}{J. Glaciology}}
  \textbf{\bibinfo{volume}{21}}, \bibinfo{pages}{55--63}
  (\bibinfo{year}{1978}).

\bibitem{Galeener1976:1474}
\bibinfo{author}{Galeener, F.~L.} \& \bibinfo{author}{Lucovsky, G.}
\newblock \bibinfo{title}{Longitudinal optical vibrations in glasses:
  {G}e{O}$_2$ and {S}i{O}$_2$}.
\newblock \emph{\bibinfo{journal}{Phys. Rev. Lett.}}
  \textbf{\bibinfo{volume}{37}}, \bibinfo{pages}{1474--1478}
  (\bibinfo{year}{1976}).

\bibitem{Restat:7277}
\bibinfo{author}{Resat, H.}, \bibinfo{author}{Raineri, F.~O.} \&
  \bibinfo{author}{Friedman, H.~L.}
\newblock \bibinfo{title}{Studies of the optical like high frequency dispersion
  mode in liquid water}.
\newblock \emph{\bibinfo{journal}{J. Chem. Phys.}}
  \textbf{\bibinfo{volume}{98}}, \bibinfo{pages}{7277} (\bibinfo{year}{1993}).

\bibitem{Kornyshev1998:1939}
\bibinfo{author}{Bopp, P.~A.}, \bibinfo{author}{Kornyshev, A.~A.} \&
  \bibinfo{author}{Sutmann, G.}
\newblock \bibinfo{title}{Frequency and wave-vector dependent dielectric
  function of water: Collective modes and relaxation spectra}.
\newblock \emph{\bibinfo{journal}{J. Chem. Phys.}}
  \textbf{\bibinfo{volume}{109}}, \bibinfo{pages}{1939} (\bibinfo{year}{1998}).

\bibitem{Ricci1989:7226}
\bibinfo{author}{Ricci, M.~A.}, \bibinfo{author}{Rocca, D.},
  \bibinfo{author}{Ruocco, G.} \& \bibinfo{author}{Vallauri, R.}
\newblock \bibinfo{title}{Theoretical and computer-simulation study of the
  density fluctuations in liquid water}.
\newblock \emph{\bibinfo{journal}{Phys. Rev. A}} \textbf{\bibinfo{volume}{40}},
  \bibinfo{pages}{7226--7238} (\bibinfo{year}{1989}).

\bibitem{Lyddane1941:673}
\bibinfo{author}{Lyddane, R.~H.}, \bibinfo{author}{Sachs, R.~G.} \&
  \bibinfo{author}{Teller, E.}
\newblock \bibinfo{title}{On the polar vibrations of alkali halides}.
\newblock \emph{\bibinfo{journal}{Phys. Rev.}} \textbf{\bibinfo{volume}{59}},
  \bibinfo{pages}{673--676} (\bibinfo{year}{1941}).

\bibitem{Barker1975:4071}
\bibinfo{author}{Barker, A.~S.}
\newblock \bibinfo{title}{Long-wavelength soft modes, central peaks, and the
  {L}yddane{-}{S}achs{-}{T}eller relation}.
\newblock \emph{\bibinfo{journal}{Phys. Rev. B}} \textbf{\bibinfo{volume}{12}},
  \bibinfo{pages}{4071--4084} (\bibinfo{year}{1975}).

\bibitem{Sievers1990:3455}
\bibinfo{author}{Sievers, A.~J.} \& \bibinfo{author}{Page, J.~B.}
\newblock \bibinfo{title}{Generalized {L}yddane-{S}achs-{T}eller relation and
  disordered solids}.
\newblock \emph{\bibinfo{journal}{Phys. Rev. B}} \textbf{\bibinfo{volume}{41}},
  \bibinfo{pages}{3455--3459} (\bibinfo{year}{1990}).

\bibitem{Payne1984:351}
\bibinfo{author}{Payne, M.} \& \bibinfo{author}{Inkson, J.}
\newblock \bibinfo{title}{Longitudinal-optic-transverse-optic vibrational mode
  splittings in tetrahedral network glasses}.
\newblock \emph{\bibinfo{journal}{Journal of Non-Crystalline Solids}}
  \textbf{\bibinfo{volume}{68}}, \bibinfo{pages}{351 -- 360}
  (\bibinfo{year}{1984}).

\bibitem{Sekimoto1982:3411}
\bibinfo{author}{Sekimoto, K.} \& \bibinfo{author}{Matsubara, T.}
\newblock \bibinfo{title}{To-lo splittings of glassy dielectrics}.
\newblock \emph{\bibinfo{journal}{Phys. Rev. B}} \textbf{\bibinfo{volume}{26}},
  \bibinfo{pages}{3411} (\bibinfo{year}{1982}).

\bibitem{Decius1977}
\bibinfo{author}{Decius, J.~C.} \& \bibinfo{author}{Hexter, R.~M.}
\newblock \emph{\bibinfo{title}{Molecular Vibrations in Crystals}}
  (\bibinfo{publisher}{McGraw-Hill, USA}, \bibinfo{year}{1977}).

\bibitem{Decius1968:1387}
\bibinfo{author}{Decius, J.~C.}
\newblock \bibinfo{title}{Dipolar coupling and molecular vibration in crystals.
  i. general theory}.
\newblock \emph{\bibinfo{journal}{J. Chem. Phys.}}
  \textbf{\bibinfo{volume}{49}}, \bibinfo{pages}{1387--1392}
  (\bibinfo{year}{1968}).

\bibitem{Burba2011:134503}
\bibinfo{author}{Burba, C.~M.} \& \bibinfo{author}{Frech, R.}
\newblock \bibinfo{title}{Existence of optical phonons in the room temperature
  ionic liquid 1-ethyl-3-methylimidazolium trifluoromethanesulfonate}.
\newblock \emph{\bibinfo{journal}{J. Chem. Phys.}}
  \textbf{\bibinfo{volume}{134}}, \bibinfo{pages}{134503}
  (\bibinfo{year}{2011}).

\bibitem{Zasetsky2005:2760}
\bibinfo{author}{Zasetsky, A.~Y.}, \bibinfo{author}{Khalizov, A.~F.},
  \bibinfo{author}{Earle, M.~E.} \& \bibinfo{author}{Sloan, J.~J.}
\newblock \bibinfo{title}{Frequency dependent complex refractive indices of
  supercooled liquid water and ice determined from aerosol extinction spectra}.
\newblock \emph{\bibinfo{journal}{The Journal of Physical Chemistry A}}
  \textbf{\bibinfo{volume}{109}}, \bibinfo{pages}{2760} (\bibinfo{year}{2005}).

\bibitem{Hale:73}
\bibinfo{author}{Hale, G.} \& \bibinfo{author}{Querry, M.}
\newblock \bibinfo{title}{Optical constants of water in the 200-nm to 200- mu m
  wavelength region}.
\newblock \emph{\bibinfo{journal}{Appl. Opt.}} \textbf{\bibinfo{volume}{12}},
  \bibinfo{pages}{555} (\bibinfo{year}{1973}).

\bibitem{Elton2014:124504}
\bibinfo{author}{Elton, D.~C.} \& \bibinfo{author}{Fern{\'a}ndez-Serra, M.-V.}
\newblock \bibinfo{title}{Polar nanoregions in water: A study of the dielectric
  properties of {TIP4P}/2005, {TIP4P}/2005f and {TTM3F}}.
\newblock \emph{\bibinfo{journal}{J. Chem. Phys.}}
  \textbf{\bibinfo{volume}{140}}, \bibinfo{pages}{124504}
  (\bibinfo{year}{2014}).

\bibitem{Madden2007:467}
\bibinfo{author}{Madden, P.} \& \bibinfo{author}{Kivelson, D.}
\newblock \bibinfo{title}{A consistent molecular treatment of dielectric
  phenomena}.
\newblock In \emph{\bibinfo{booktitle}{Adv. Chem. Phys.}}, \bibinfo{pages}{467}
  (\bibinfo{publisher}{John Wiley \& Sons, Inc.}, \bibinfo{year}{2007}).

\bibitem{Hansen2006:341}
\bibinfo{author}{Hansen, J.-P.} \& \bibinfo{author}{McDonald, I.~R.}
\newblock \bibinfo{title}{Chapter 11 - molecular liquids}.
\newblock In \bibinfo{editor}{Hansen, J.-P.} \& \bibinfo{editor}{McDonald,
  I.~R.} (eds.) \emph{\bibinfo{booktitle}{Theory of Simple Liquids}},
  \bibinfo{pages}{341} (\bibinfo{publisher}{Academic Press},
  \bibinfo{year}{2006}), \bibinfo{edition}{3rd} edn.

\bibitem{Restat1992:2618}
\bibinfo{author}{Resat, H.}, \bibinfo{author}{Raineri, F.~O.} \&
  \bibinfo{author}{Friedman, H.~L.}
\newblock \bibinfo{title}{A dielectric theory of the optical‐like
  high‐frequency mode in liquid water}.
\newblock \emph{\bibinfo{journal}{J. Chem. Phys.}}
  \textbf{\bibinfo{volume}{97}}, \bibinfo{pages}{2618} (\bibinfo{year}{1992}).

\bibitem{Heyden2012:2135}
\bibinfo{author}{Heyden, M.} \emph{et~al.}
\newblock \bibinfo{title}{Understanding the origins of dipolar couplings and
  correlated motion in the vibrational spectrum of water}.
\newblock \emph{\bibinfo{journal}{J. Phys. Chem. Lett.}}
  \textbf{\bibinfo{volume}{3}}, \bibinfo{pages}{2135--2140}
  (\bibinfo{year}{2012}).

\bibitem{woodward1972}
\bibinfo{author}{Woodward, L.}
\newblock \emph{\bibinfo{title}{Introduction to the theory of molecular
  vibrations and vibrational spectroscopy}} (\bibinfo{publisher}{Clarendon
  Press}, \bibinfo{year}{1972}).

\bibitem{Wagner2005:7099}
\bibinfo{author}{Wagner, R.} \emph{et~al.}
\newblock \bibinfo{title}{Mid-infrared extinction spectra and optical constants
  of supercooled water droplets}.
\newblock \emph{\bibinfo{journal}{J. Phys. Chem. A}}
  \textbf{\bibinfo{volume}{109}}, \bibinfo{pages}{7099--7112}
  (\bibinfo{year}{2005}).

\bibitem{Vinh2015:164502}
\bibinfo{author}{Vinh, N.~Q.} \emph{et~al.}
\newblock \bibinfo{title}{High-precision gigahertz-to-terahertz spectroscopy of
  aqueous salt solutions as a probe of the femtosecond-to-picosecond dynamics
  of liquid water}.
\newblock \emph{\bibinfo{journal}{J. Chem. Phys.}}
  \textbf{\bibinfo{volume}{142}}.

\bibitem{Ellison2007:1}
\bibinfo{author}{Ellison, W.~J.}
\newblock \bibinfo{title}{Permittivity of pure water, at standard atmospheric
  pressure, over the frequency range {0-25} {TH}z and the temperature range
  0–100°c}.
\newblock \emph{\bibinfo{journal}{J. Phys. Chem. Ref.D at.}}
  \textbf{\bibinfo{volume}{36}}, \bibinfo{pages}{1--18} (\bibinfo{year}{2007}).

\bibitem{Iwano2010:063601}
\bibinfo{author}{Iwano, K.}, \bibinfo{author}{Yokoo, T.},
  \bibinfo{author}{Oguro, M.} \& \bibinfo{author}{Ikeda, S.}
\newblock \bibinfo{title}{Propagating librations in ice xi: Model analysis and
  coherent inelastic neutron scattering experiment}.
\newblock \emph{\bibinfo{journal}{Journal of the Physical Society of Japan}}
  \textbf{\bibinfo{volume}{79}}, \bibinfo{pages}{063601}
  (\bibinfo{year}{2010}).

\bibitem{Shigenari2012:174504}
\bibinfo{author}{Shigenari, T.} \& \bibinfo{author}{Abe, K.}
\newblock \bibinfo{title}{Vibrational modes of hydrogens in the proton ordered
  phase {XI} of ice: {R}aman spectra above 400 cm$^{−1}$}.
\newblock \emph{\bibinfo{journal}{J. Chem. Phys.}}
  \textbf{\bibinfo{volume}{136}} (\bibinfo{year}{2012}).

\bibitem{Wan2013:4124}
\bibinfo{author}{Wan, Q.}, \bibinfo{author}{Spanu, L.}, \bibinfo{author}{Galli,
  G.~A.} \& \bibinfo{author}{Gygi, F.}
\newblock \bibinfo{title}{Raman spectra of liquid water from ab initio
  molecular dynamics: Vibrational signatures of charge fluctuations in the
  hydrogen bond network}.
\newblock \emph{\bibinfo{journal}{Journal of Chemical Theory and Computation}}
  \textbf{\bibinfo{volume}{9}}, \bibinfo{pages}{4124--4130}
  (\bibinfo{year}{2013}).

\bibitem{Lobo1973:5992}
\bibinfo{author}{Lobo, R.}, \bibinfo{author}{Robinson, J.~E.} \&
  \bibinfo{author}{Rodriguez, S.}
\newblock \bibinfo{title}{High frequency dielectric response of dipolar
  liquids}.
\newblock \emph{\bibinfo{journal}{J. Chem. Phys.}}
  \textbf{\bibinfo{volume}{59}}, \bibinfo{pages}{5992--6008}
  (\bibinfo{year}{1973}).

\bibitem{Pollock1981:950}
\bibinfo{author}{Pollock, E.~L.} \& \bibinfo{author}{Alder, B.~J.}
\newblock \bibinfo{title}{Frequency-dependent dielectric response in polar
  liquids}.
\newblock \emph{\bibinfo{journal}{Phys. Rev. Lett.}}
  \textbf{\bibinfo{volume}{46}}, \bibinfo{pages}{950--953}
  (\bibinfo{year}{1981}).

\bibitem{Chandra1990:6833}
\bibinfo{author}{Chandra, A.} \& \bibinfo{author}{Bagchi, B.}
\newblock \bibinfo{title}{Collective excitations in a dense dipolar liquid: How
  important are dipolarons in the polarization relaxation of common dipolar
  liquids?}
\newblock \emph{\bibinfo{journal}{J. Chem. Phys.}}
  \textbf{\bibinfo{volume}{92}}, \bibinfo{pages}{6833} (\bibinfo{year}{1990}).

\bibitem{Haughhey1987:4934}
\bibinfo{author}{Haughney, M.}, \bibinfo{author}{Ferrario, M.} \&
  \bibinfo{author}{McDonald, I.~R.}
\newblock \bibinfo{title}{Molecular-dynamics simulation of liquid methanol}.
\newblock \emph{\bibinfo{journal}{The Journal of Physical Chemistry}}
  \textbf{\bibinfo{volume}{91}}, \bibinfo{pages}{4934--4940}
  (\bibinfo{year}{1987}).

\bibitem{Yamaguchi2000:8976}
\bibinfo{author}{Yamaguchi, T.}, \bibinfo{author}{Benmore, C.~J.} \&
  \bibinfo{author}{Soper, A.~K.}
\newblock \bibinfo{title}{The structure of subcritical and supercritical
  methanol by neutron diffraction, empirical potential structure refinement,
  and spherical harmonic analysis}.
\newblock \emph{\bibinfo{journal}{J. Chem. Phys.}}
  \textbf{\bibinfo{volume}{112}}, \bibinfo{pages}{8976--8987}
  (\bibinfo{year}{2000}).

\bibitem{Matsumoto1990:1981}
\bibinfo{author}{Matsumoto, M.} \& \bibinfo{author}{Gubbins, K.~E.}
\newblock \bibinfo{title}{Hydrogen bonding in liquid methanol}.
\newblock \emph{\bibinfo{journal}{J. Chem. Phys.}}
  \textbf{\bibinfo{volume}{93}}, \bibinfo{pages}{1981--1994}
  (\bibinfo{year}{1990}).

\bibitem{Crowder1967:914}
\bibinfo{author}{Crowder, G.~A.} \& \bibinfo{author}{Cook, B.~R.}
\newblock \bibinfo{title}{Acetonitrile: far-infrared spectra and chemical
  thermodynamic properties. discussion of an entropy discrepancy}.
\newblock \emph{\bibinfo{journal}{The Journal of Physical Chemistry}}
  \textbf{\bibinfo{volume}{71}}, \bibinfo{pages}{914--916}
  (\bibinfo{year}{1967}).

\bibitem{Santucci2006:225701}
\bibinfo{author}{Santucci, S.~C.}, \bibinfo{author}{Fioretto, D.},
  \bibinfo{author}{Comez, L.}, \bibinfo{author}{Gessini, A.} \&
  \bibinfo{author}{Masciovecchio, C.}
\newblock \bibinfo{title}{Is there any fast sound in water?}
\newblock \emph{\bibinfo{journal}{Phys. Rev. Lett.}}
  \textbf{\bibinfo{volume}{97}}, \bibinfo{pages}{225701}
  (\bibinfo{year}{2006}).

\bibitem{Sampoli1997:1678}
\bibinfo{author}{Sampoli, M.}, \bibinfo{author}{Ruocco, G.} \&
  \bibinfo{author}{Sette, F.}
\newblock \bibinfo{title}{Mixing of longitudinal and transverse dynamics in
  liquid water}.
\newblock \emph{\bibinfo{journal}{Phys. Rev. Lett.}}
  \textbf{\bibinfo{volume}{79}}, \bibinfo{pages}{1678--1681}
  (\bibinfo{year}{1997}).

\bibitem{Cho1994:6672}
\bibinfo{author}{Cho, M.}, \bibinfo{author}{Fleming, G.~R.},
  \bibinfo{author}{Saito, S.}, \bibinfo{author}{Ohmine, I.} \&
  \bibinfo{author}{Stratt, R.~M.}
\newblock \bibinfo{title}{Instantaneous normal mode analysis of liquid water}.
\newblock \emph{\bibinfo{journal}{J. Chem. Phys.}}
  \textbf{\bibinfo{volume}{100}}, \bibinfo{pages}{6672--6683}
  (\bibinfo{year}{1994}).

\bibitem{Garberoglio2002:3278}
\bibinfo{author}{Garberoglio, G.}, \bibinfo{author}{Vallauri, R.} \&
  \bibinfo{author}{Sutmann, G.}
\newblock \bibinfo{title}{Instantaneous normal mode analysis of correlated
  cluster motions in hydrogen bonded liquids}.
\newblock \emph{\bibinfo{journal}{J. Chem. Phys.}}
  \textbf{\bibinfo{volume}{117}}, \bibinfo{pages}{3278--3288}
  (\bibinfo{year}{2002}).

\bibitem{Seldmeier2014:054512}
\bibinfo{author}{Sedlmeier, F.}, \bibinfo{author}{Shadkhoo, S.},
  \bibinfo{author}{Bruinsma, R.} \& \bibinfo{author}{Netz, R.~R.}
\newblock \bibinfo{title}{Charge/mass dynamic structure factors of water and
  applications to dielectric friction and electroacoustic conversion}.
\newblock \emph{\bibinfo{journal}{J. Chem. Phys.}}
  \textbf{\bibinfo{volume}{140}}, \bibinfo{pages}{054512}
  (\bibinfo{year}{2014}).

\bibitem{Bosma1993:4413}
\bibinfo{author}{Bosma, W.~B.}, \bibinfo{author}{Fried, L.~E.} \&
  \bibinfo{author}{Mukamel, S.}
\newblock \bibinfo{title}{Simulation of the intermolecular vibrational spectra
  of liquid water and water clusters}.
\newblock \emph{\bibinfo{journal}{J. Chem. Phys.}}
  \textbf{\bibinfo{volume}{98}}, \bibinfo{pages}{4413--4421}
  (\bibinfo{year}{1993}).

\bibitem{Nibali2014:12800}
\bibinfo{author}{Conti~Nibali, V.} \& \bibinfo{author}{Havenith, M.}
\newblock \bibinfo{title}{New insights into the role of water in biological
  function: Studying solvated biomolecules using terahertz absorption
  spectroscopy in conjunction with molecular dynamics simulations}.
\newblock \emph{\bibinfo{journal}{J. Am. Chem. Soc.}}
  \textbf{\bibinfo{volume}{136}}, \bibinfo{pages}{12800}
  (\bibinfo{year}{2014}).

\bibitem{Ebbinghaus2007:20749}
\bibinfo{author}{Ebbinghaus, S.} \emph{et~al.}
\newblock \bibinfo{title}{An extended dynamical hydration shell around
  proteins}.
\newblock \emph{\bibinfo{journal}{Proc. Natl. Acad. Sci. USA}}
  \textbf{\bibinfo{volume}{104}}, \bibinfo{pages}{20749}
  (\bibinfo{year}{2007}).

\bibitem{Kim2008:6486}
\bibinfo{author}{Kim, S.}, \bibinfo{author}{Born, B.},
  \bibinfo{author}{Havenith, M.} \& \bibinfo{author}{Gruebele, M.}
\newblock \bibinfo{title}{Real-time detection of protein-water dynamics upon
  protein folding by terahertz absorption spectroscopy}.
\newblock \emph{\bibinfo{journal}{Angewandte Chemie International Edition}}
  \textbf{\bibinfo{volume}{47}}, \bibinfo{pages}{6486} (\bibinfo{year}{2008}).

\bibitem{Kubo1957:570}
\bibinfo{author}{Kubo, R.}
\newblock \bibinfo{title}{Statistical-mechanical theory of irreversible
  processes. i. general theory and simple applications to magnetic and
  conduction problems}.
\newblock \emph{\bibinfo{journal}{J. Phys. Soc. Jap.}}
  \textbf{\bibinfo{volume}{12}}, \bibinfo{pages}{570--586}
  (\bibinfo{year}{1957}).

\bibitem{Raineri1993:8910}
\bibinfo{author}{Raineri, F.~O.} \& \bibinfo{author}{Friedman, H.~L.}
\newblock \bibinfo{title}{Static transverse dielectric function of model
  molecular fluids}.
\newblock \emph{\bibinfo{journal}{J. Chem. Phys.}}
  \textbf{\bibinfo{volume}{98}}, \bibinfo{pages}{8910--8918}
  (\bibinfo{year}{1993}).

\bibitem{Fuentes-Azcatl2014:1263}
\bibinfo{author}{Fuentes-Azcatl, R.} \& \bibinfo{author}{Alejandre, J.}
\newblock \bibinfo{title}{Non-polarizable force field of water based on the
  dielectric constant: {TIP4P}/$\varepsilon$}.
\newblock \emph{\bibinfo{journal}{J. Phys. Chem. B}}
  \textbf{\bibinfo{volume}{118}}, \bibinfo{pages}{1263--1272}
  (\bibinfo{year}{2014}).

\bibitem{gonzalez:224516}
\bibinfo{author}{Gonzalez, M.~A.} \& \bibinfo{author}{Abascal, J. L.~F.}
\newblock \bibinfo{title}{A flexible model for water based on {TIP4P}/2005}.
\newblock \emph{\bibinfo{journal}{J. Chem. Phys.}}
  \textbf{\bibinfo{volume}{135}}, \bibinfo{pages}{224516}
  (\bibinfo{year}{2011}).

\bibitem{fanourgakis:074506}
\bibinfo{author}{Fanourgakis, G.~S.} \& \bibinfo{author}{Xantheas, S.~S.}
\newblock \bibinfo{title}{Development of transferable interaction potentials
  for water. v. extension of the flexible, polarizable, thole-type model
  potential ({TTM}3-{F}, v. 3.0) to describe the vibrational spectra of water
  clusters and liquid water}.
\newblock \emph{\bibinfo{journal}{J. Chem. Phys.}}
  \textbf{\bibinfo{volume}{128}}, \bibinfo{pages}{074506}
  (\bibinfo{year}{2008}).

\bibitem{Wang2004:1157}
\bibinfo{author}{Wang, J.} \emph{et~al.}
\newblock \bibinfo{title}{Development and testing of a general {AMBER} force
  field}.
\newblock \emph{\bibinfo{journal}{J. Compt. Chem.}}
  \textbf{\bibinfo{volume}{25}}, \bibinfo{pages}{1157} (\bibinfo{year}{2004}).

\bibitem{Caleman2012:61}
\bibinfo{author}{Caleman, C.} \emph{et~al.}
\newblock \bibinfo{title}{Force field benchmark of organic liquids: Density,
  enthalpy of vaporization, heat capacities, surface tension, isothermal
  compressibility, volumetric expansion coefficient, and dielectric constant}.
\newblock \emph{\bibinfo{journal}{J. Chem. Theo. Comp.}}
  \textbf{\bibinfo{volume}{8}}, \bibinfo{pages}{61} (\bibinfo{year}{2012}).

\bibitem{Hess:435}
\bibinfo{author}{Hess, B.}, \bibinfo{author}{Kutzner, C.},
  \bibinfo{author}{van~der Spoel, D.} \& \bibinfo{author}{Lindahl, E.}
\newblock \bibinfo{title}{{GROMACS} 4: Algorithms for highly efficient,
  load-balanced, and scalable molecular simulation}.
\newblock \emph{\bibinfo{journal}{J Chem. Theo. Comp.}}
  \textbf{\bibinfo{volume}{4}}, \bibinfo{pages}{435--447}
  (\bibinfo{year}{2008}).

\bibitem{toda1991}
\bibinfo{author}{Toda, M.}, \bibinfo{author}{Kubo, R.},
  \bibinfo{author}{Sait{\=o}, N.} \& \bibinfo{author}{Hashitsume, N.}
\newblock \emph{\bibinfo{title}{Statistical Phys. II: Nonequilibrium
  Statistical Mechanics}}.
\newblock Series C (\bibinfo{publisher}{Springer Berlin Heidelberg},
  \bibinfo{year}{1991}).

\bibitem{VanVleck1945:227}
\bibinfo{author}{Van~Vleck, J.~H.} \& \bibinfo{author}{Weisskopf, V.~F.}
\newblock \bibinfo{title}{On the shape of collision-broadened lines}.
\newblock \emph{\bibinfo{journal}{Rev. Mod. Phys.}}
  \textbf{\bibinfo{volume}{17}}, \bibinfo{pages}{227--236}
  (\bibinfo{year}{1945}).

\bibitem{Buchner1999:57}
\bibinfo{author}{Buchner, R.}, \bibinfo{author}{Barthel, J.} \&
  \bibinfo{author}{Stauber, J.}
\newblock \bibinfo{title}{The dielectric relaxation of water between 0°c and
  35°c}.
\newblock \emph{\bibinfo{journal}{Chem. Phys. Lett.}}
  \textbf{\bibinfo{volume}{306}}, \bibinfo{pages}{57} (\bibinfo{year}{1999}).

\bibitem{Bolla1933:101}
\bibinfo{author}{Bolla, G.}
\newblock \bibinfo{title}{Su alcune nuove bande {R}aman dell’acqua}.
\newblock \emph{\bibinfo{journal}{Il Nuovo Cimento}}
  \textbf{\bibinfo{volume}{10}}, \bibinfo{pages}{101--107}
  (\bibinfo{year}{1933}).

\bibitem{Walrafen1962:1035}
\bibinfo{author}{Walrafen, G.~E.}
\newblock \bibinfo{title}{Raman spectral studies of the effects of electrolytes
  on water}.
\newblock \emph{\bibinfo{journal}{J. Chem. Phys.}}
  \textbf{\bibinfo{volume}{36}}, \bibinfo{pages}{1035--1042}
  (\bibinfo{year}{1962}).

\bibitem{Heyden2010:12068}
\bibinfo{author}{Heyden, M.} \emph{et~al.}
\newblock \bibinfo{title}{Dissecting the {TH}z spectrum of liquid water from
  first principles via correlations in time and space}.
\newblock \emph{\bibinfo{journal}{Proc. Natl. Acad. Sci. USA}}
  \textbf{\bibinfo{volume}{107}}, \bibinfo{pages}{12068--12073}
  (\bibinfo{year}{2010}).

\bibitem{Bertolini:1065}
\bibinfo{author}{Bertolini, D.} \& \bibinfo{author}{Tani, A.}
\newblock \bibinfo{title}{The frequency and wavelength dependent dielectric
  permittivity of water}.
\newblock \emph{\bibinfo{journal}{Mol. Phys.}} \textbf{\bibinfo{volume}{75}},
  \bibinfo{pages}{1065--1088} (\bibinfo{year}{1992}).

\bibitem{Bagchi1993:133}
\bibinfo{author}{Bagchi, B.} \& \bibinfo{author}{Chandra, A.}
\newblock \bibinfo{title}{Molecular theory of underdamped dielectric
  relaxation: understanding collective effects in dipolar liquids}.
\newblock \emph{\bibinfo{journal}{Chem. Phys.}} \textbf{\bibinfo{volume}{173}},
  \bibinfo{pages}{133 -- 141} (\bibinfo{year}{1993}).

\bibitem{ferraro1978}
\bibinfo{author}{Ferraro, R.} \& \bibinfo{author}{Basile, L.}
\newblock \emph{\bibinfo{title}{Fourier Transform Infrared Spectra:
  Applications to Chem. Systems}}, vol. \bibinfo{volume}{v. 1}
  (\bibinfo{publisher}{Elsevier Science}, \bibinfo{year}{1978}).

\bibitem{Bagchi1992:5126}
\bibinfo{author}{Bagchi, B.} \& \bibinfo{author}{Chandra, A.}
\newblock \bibinfo{title}{Ultrafast solvation dynamics: Molecular explanation
  of computer simulation results in a simple dipolar solvent}.
\newblock \emph{\bibinfo{journal}{J. Chem. Phys.}}
  \textbf{\bibinfo{volume}{97}} (\bibinfo{year}{1992}).

\bibitem{Hill1971:2322}
\bibinfo{author}{Hill, N.~E.}
\newblock \bibinfo{title}{The influence of the poley absorption on the inertial
  fall-off of the dielectric absorption}.
\newblock \emph{\bibinfo{journal}{J. Phys. C: Solid State Phys.}}
  \textbf{\bibinfo{volume}{4}}, \bibinfo{pages}{2322} (\bibinfo{year}{1971}).

\bibitem{H69}
\bibinfo{author}{Nora E.~Hill, A. P. M.~D., W. E.~Vaughan}.
\newblock \emph{\bibinfo{title}{Dielectric Properties and Molecular Behaviour}}
  (\bibinfo{publisher}{Van Norstrand Reinhold Company}, \bibinfo{address}{New
  York}, \bibinfo{year}{1963}).

\bibitem{Poley1955:337}
\bibinfo{author}{Poley, J.}
\newblock \bibinfo{title}{Microwave dispersion of some polar liquids}.
\newblock \emph{\bibinfo{journal}{App. Sci. Res. B}}
  \textbf{\bibinfo{volume}{4}}, \bibinfo{pages}{337--387}
  (\bibinfo{year}{1955}).

\end{thebibliography}
%\bibliographystyle{naturemag}

%% For Tables, put caption above table
%%
%% Table caption should start with a capital letter, continue with lower case
%% and not have a period at the end
%% Using @{\vrule height ?? depth ?? width0pt} in the tabular preamble will
%% keep that much space between every line in the table.
%% \begin{table}
%% \caption{Repeat length of longer allele by age of onset class}
%% \begin{tabular}{@{\vrule height 10.5pt depth4pt  width0pt}lrcccc}

%----------------------------------------------------------------------------------------
%	SI
%----------------------------------------------------------------------------------------
\onecolumngrid
\newpage
\FloatBarrier
\setcounter{figure}{0}    
\setcounter{table}{0}    

\section{Supplementary information}
\begin{table}
     \begin{tabular}{c c c c c}    
type &  $\omega_{\ff{L1}}$& $\omega_{\ff{L2}}$ & $\omega_{\ff{L3}}$ & ref.   \\
  \hline   
\multirow{8}{*}{Raman}\label{ExptLibPeaks}
     & 510 & --- & 780 & Bolla (1933) \cite{Bolla1933:101} \\
     & 450 & --- & 780 & Walrafen (1962) \cite{Walrafen1962:1035} \\
     & 400 & --- & 700 & Fukasawa, et. al. (2005)\cite{Fukasawa2005:197802} \\
     & 430 & 650 & 795 & Carey, et. al. (1998)\cite{Carey1998:2669} \\
     & 440 & 540 & 770 & Castner, et. al. (1995)\cite{Castner1995:653} \\
     & 450 & 550 & 725 & Walrafen (1990)\cite{Walrafen1990:2237}\\
     & 424 & 550 & 725 & Walrafen (1986)\cite{Walrafen1986:6970} \\
     & 439 & 538 & 717 & Walrafen (1967)\cite{Walrafen1967:114} \\
infrared 
     & 380 & 665 & --- & Zelsmann (1995) \cite{Zelsmann1995:95} \\
dielectric 
     & 420 & 620 & --- & Fukasawa, et. al. (2005)\cite{Fukasawa2005:197802}  
     \end{tabular}
      \\
      \begin{center}
       {\bf Supplementary Table 1: Experimental peaks in Raman, dielectric, and IR spectra.} This table shows the correspondence between 3 peak Raman fits and 2 peak dielectric/IR fits to the librational region at 298 K.
       \end{center}
\end{table}
  
\vspace{2cm}

\begin{figure}[!h]
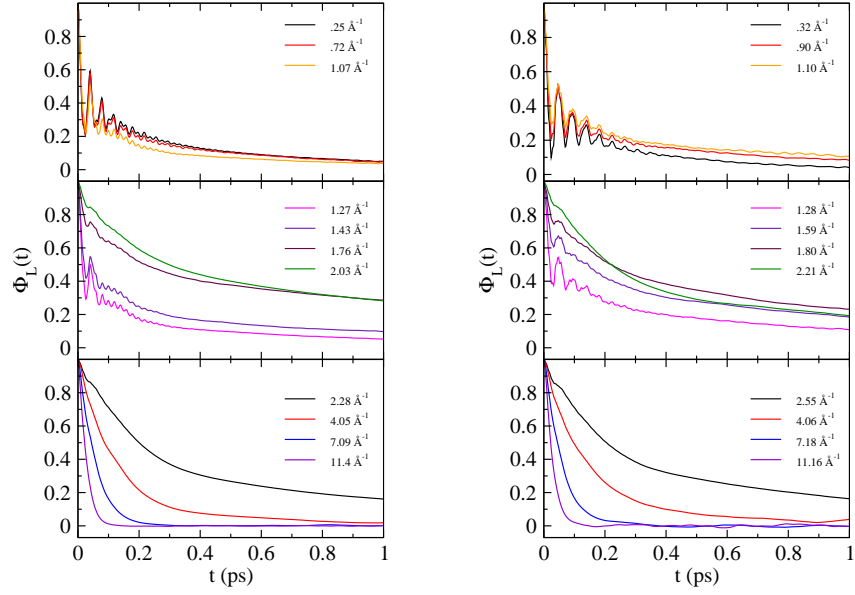

 \includegraphics[width=5cm]{phi_vs_t_flexible.eps}
 $\quad\quad\quad$
 \includegraphics[width=5cm]{phi_vs_t_TTM3F.eps}
  \caption{{\bf Longitudinal polarization relaxation functions.} Shown for 512 TIP4P2005/f (left) and TTM3F (right) at 300 K.}
\end{figure}
\begin{figure}[!h]
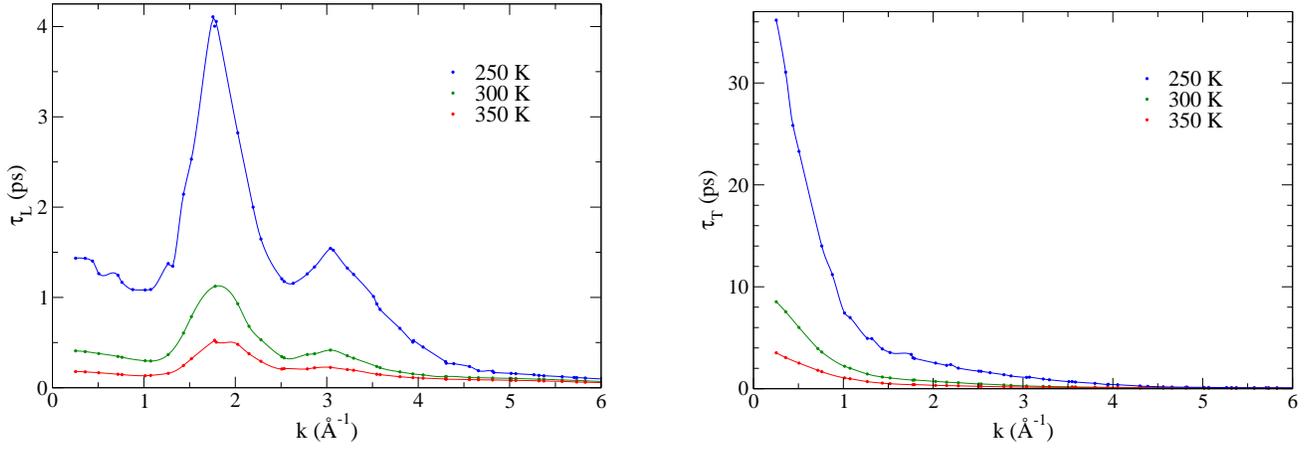

  \centering
  \includegraphics[width=8cm]{tau_vs_k_L_TIP4P2005f_512_300.eps}
   $\quad\quad\quad$
  \includegraphics[width=8cm]{tau_vs_k_T_TIP4P2005f_512_300.eps}
  \caption{{\bf Longitudinal (left) and transverse (right) relaxation times for 512 TIP4P/2005f.} Computed for the underlying exponential of the relaxation. The points are interpolated by Akima splines. The transverse relaxation time here is equal to the Debye relaxation time, which at $k = 0$ is $\approx$ 11 ps at 300 K for TIP4P/2005f. Experimentally it is 8.5 ps.\cite{Buchner1999:57} } 
  \label{realaxationL}
\end{figure}
%\begin{figure}[!h]
%  \centering   
%  \includegraphics[width=6cm]{phi_zoomed_in_TIP4Peps_512_300.eps}
%  \caption{{\bf Fine features of the longitudinal polarization correlation function for 512 TIP4P/$\varepsilon$ at 300 K.} Coherent small-magnitude %oscillations appear to persist for longer than 1 ps.}
% \label{finefeatures}
%\end{figure}
%Relaxation times can be computed by fitting exponentials to the correlation functions. Figure \ref{realaxationL} shows the longitudinal and transverse relaxation times of TIP4P/2005f at 300 K. We see that the longitudinal relaxation exhibits a peak at $k \approx 2\Ang^{-1}, \lambda = 2.1 \Ang$) and a secondary peak at $k \approx 3 \Ang^{-1}, \lambda \approx 3.1 \Ang $. The first peak is close to $k$ values corresponding to the O-H distance in the hydrogen bond, while the second is close to the O-O distance. 
\FloatBarrier
%------------------------------------------------------------------------
\subsection{Supplementary info: Dispersion relations and dampening factors}
\begin{figure}
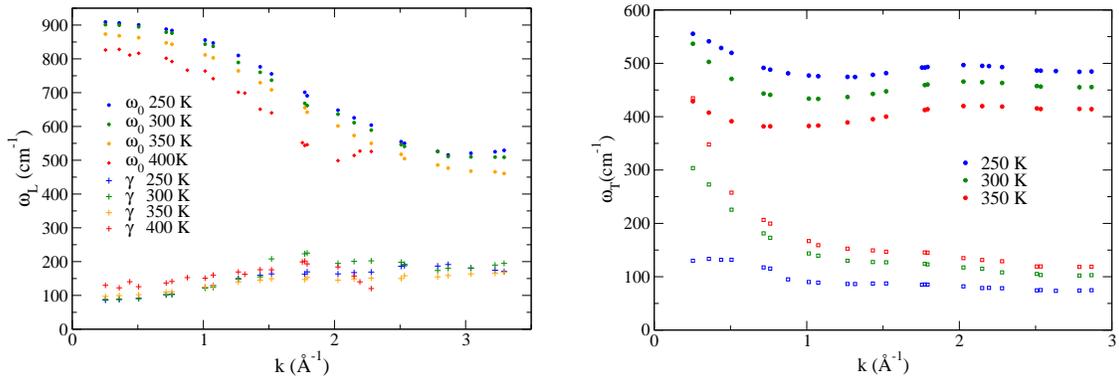

 \includegraphics[width=7cm]{w_vs_k_diff_temps.eps}$\quad\quad$
 \includegraphics[width=7cm]{w_vs_k_T.eps}
  \caption{{\bf Longitudinal (left) and transverse (right) dispersion relations (circles) and dampening factors (squares) for 512 TIP4P/2005f.} These curves were obtained from a two peak (Debye + resonant) fit. In contrast to the longitudinal mode, the transverse mode is much more damped. However, the dampening factor changes significantly with temperature, also in contrast to the longitudinal case, and at 250 K becomes relatively small. Beyond 2 $\Ang^{-1}$ the peak due to the damped resonance starts to disappear so values beyond 3 $\Ang^{-1}$ are not shown.}
\end{figure}
\FloatBarrier
 
\section{ Supplementary info:  Methanol \& acetonitrile}
\begin{figure}
\centering 
\includegraphics[width=8cm]{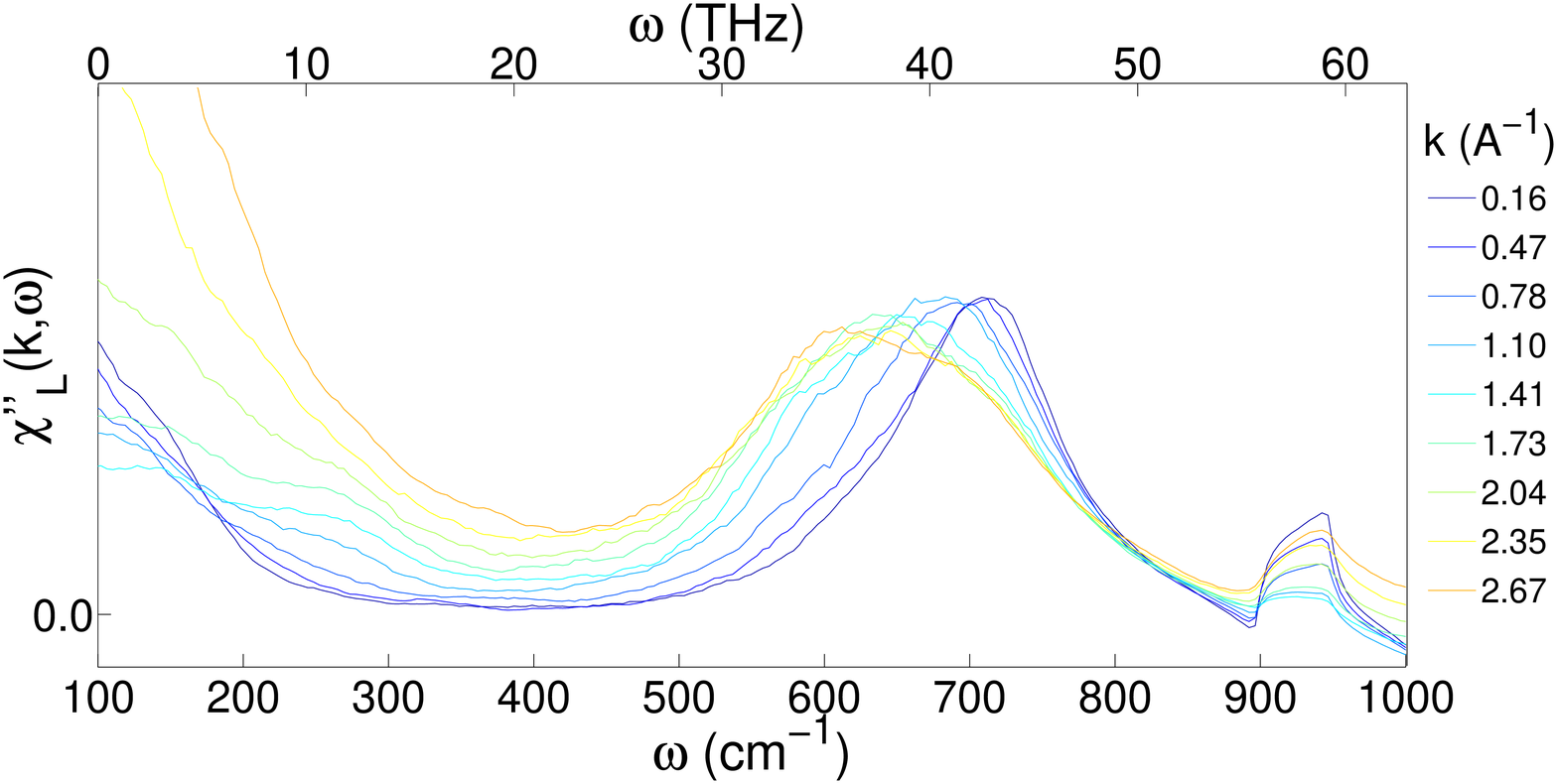}
\includegraphics[width=8cm]{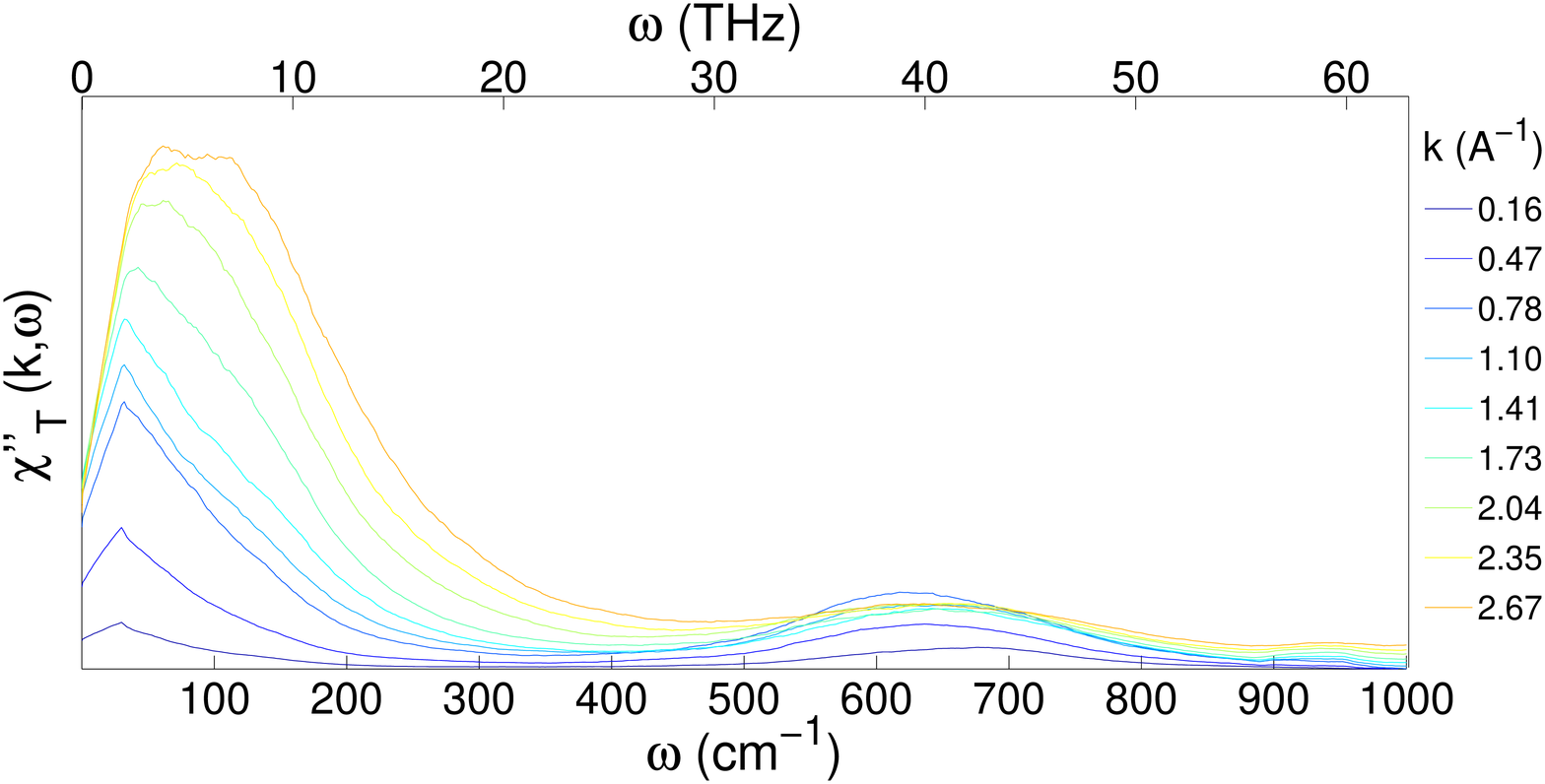}
  \caption{{\bf Longitudinal (left) and transverse (right) dielectric susceptibility for a system of 1,000 MeOH molecules.} The longitudinal librational peak at $\approx$ 700 cm$^{-1}$ clearly disperses with $k$, while the transverse peak at $\approx$ 600 cm$^{-1}$ disperses slightly with $k$. The higher frequency peaks exhibit no dispersion. The static dielectric function $\varepsilon(k,0)$ has not converged properly in the transverse case, so the magnitude of the peaks is not converged.
}\label{methanol}

\centering  
\includegraphics[width=8cm]{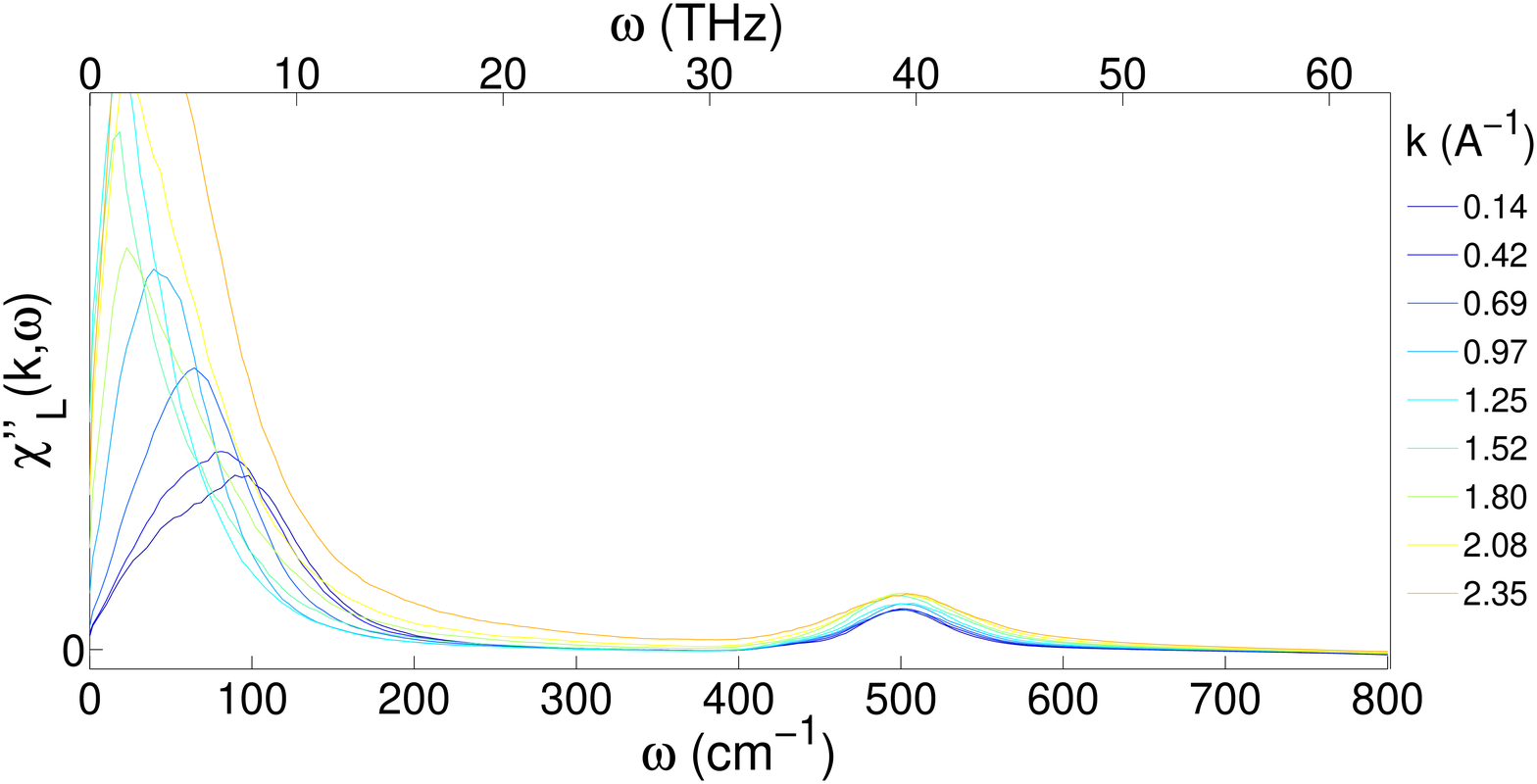}
\includegraphics[width=8cm]{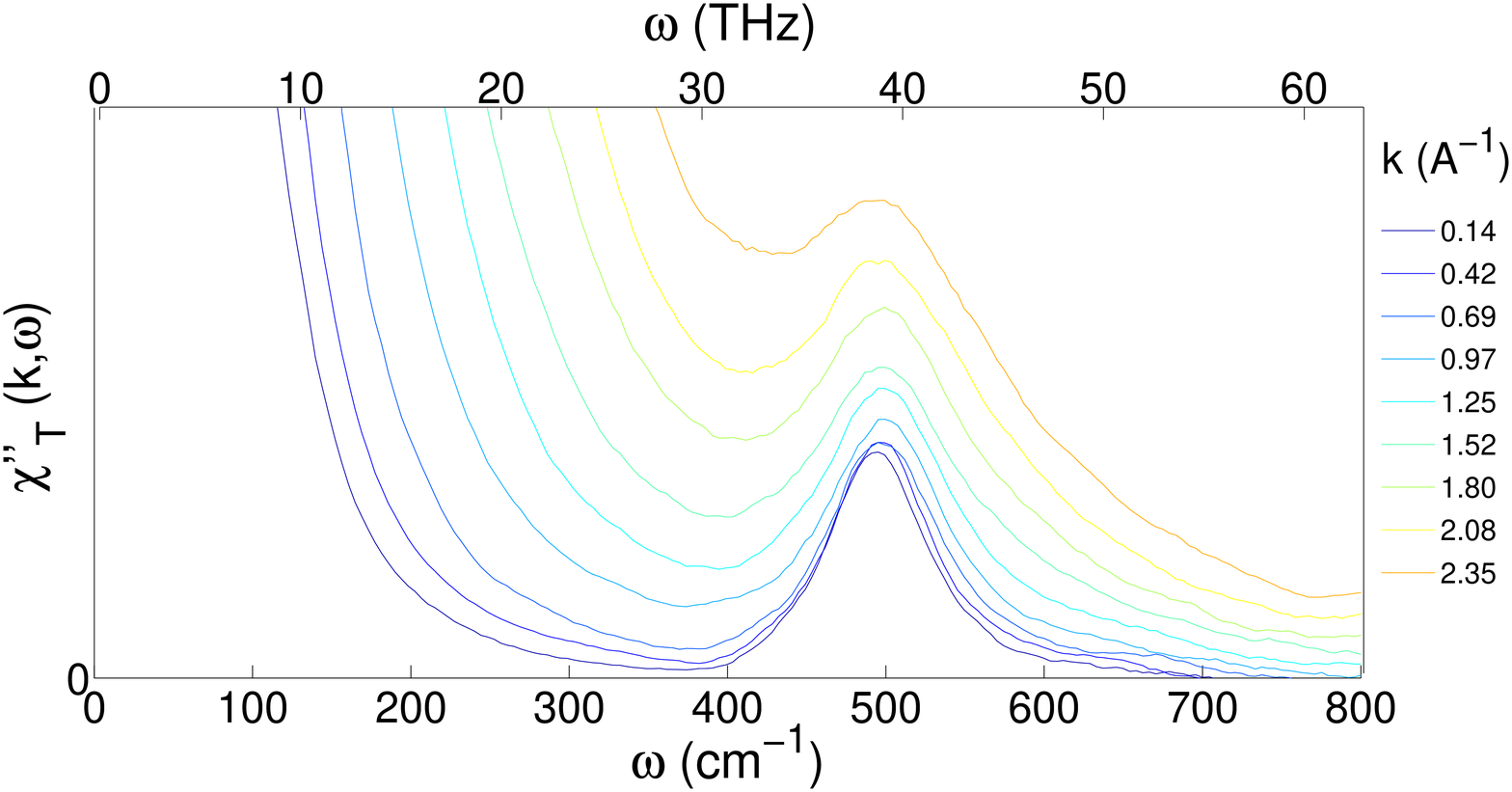}
  \caption{{\bf Longitudinal (left) and transverse (right) dielectric susceptibility for a system of 1,000 acetonitrile molecules.} The broad band which peaks at $100$ cm$^{-1}$ exhibits dispersion. We hypothesize this dispersion is due entirely to the translational modes, however we cannot say for sure since the librational and translational modes overlap in this region. The peak at $\approx$ 500 cm$^{-1}$ is due to CCN bending. The static dielectric function $\varepsilon(k,0)$ has not converged properly, so the magnitude of the transverse peaks is not converged correctly, but the position of the peaks and dispersion can be seen.}
\label{ACN}
\end{figure}
\FloatBarrier

\subsection{Supplementary info:  Examples of fitting}
\begin{figure}
\centering   \includegraphics[width=6cm]{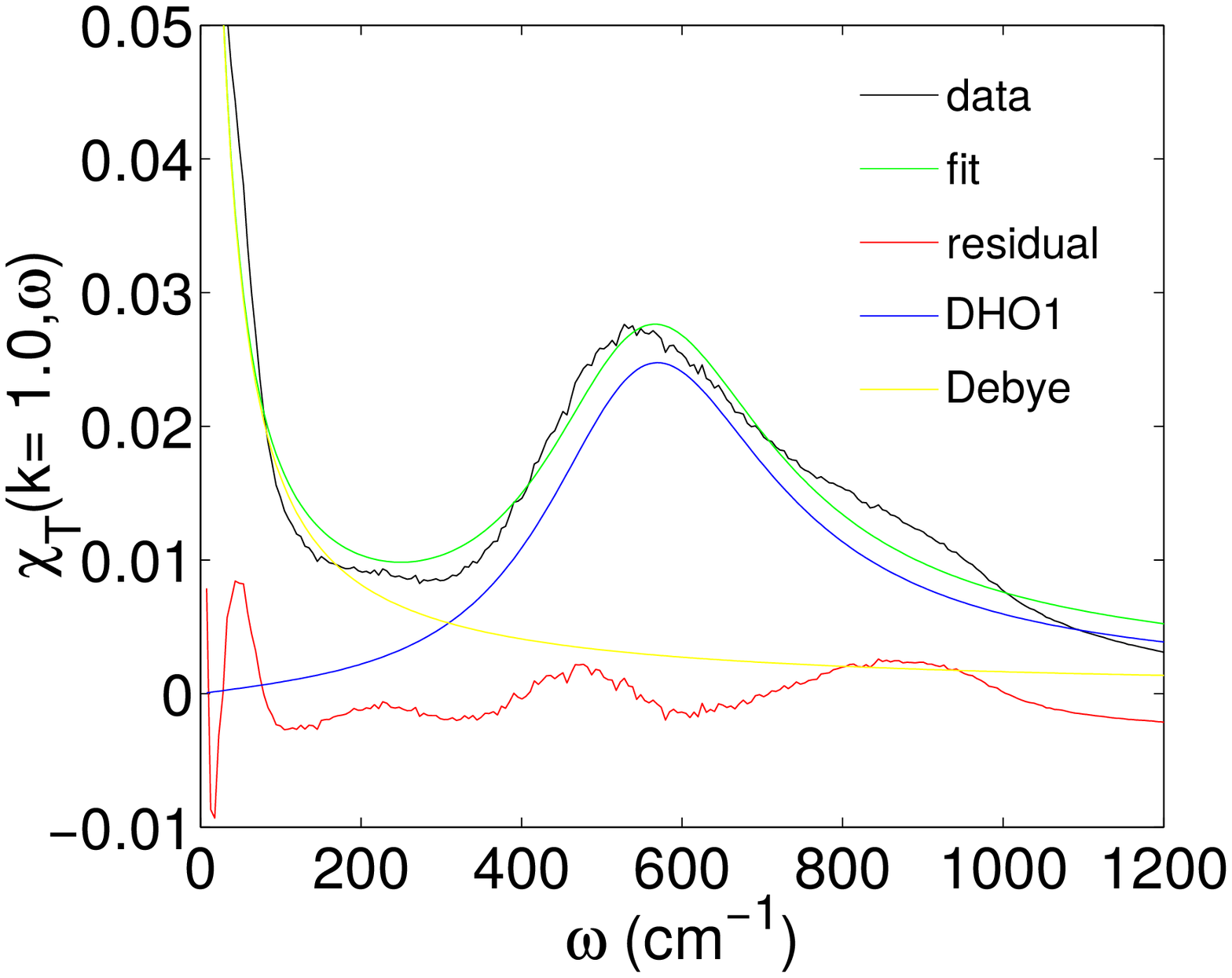}
  \caption{{\bf Example fits of the transverse susceptibility of TIP4P/2005f at 300 K.} Fit with a Debye function and one damped harmonic oscillator at $k = .25 \Ang^{-1}$ and $k = 1.4 \Ang^{-1}$. The residual show the parts not captured by the fit.}
  \centering 
   \includegraphics[width=6cm]{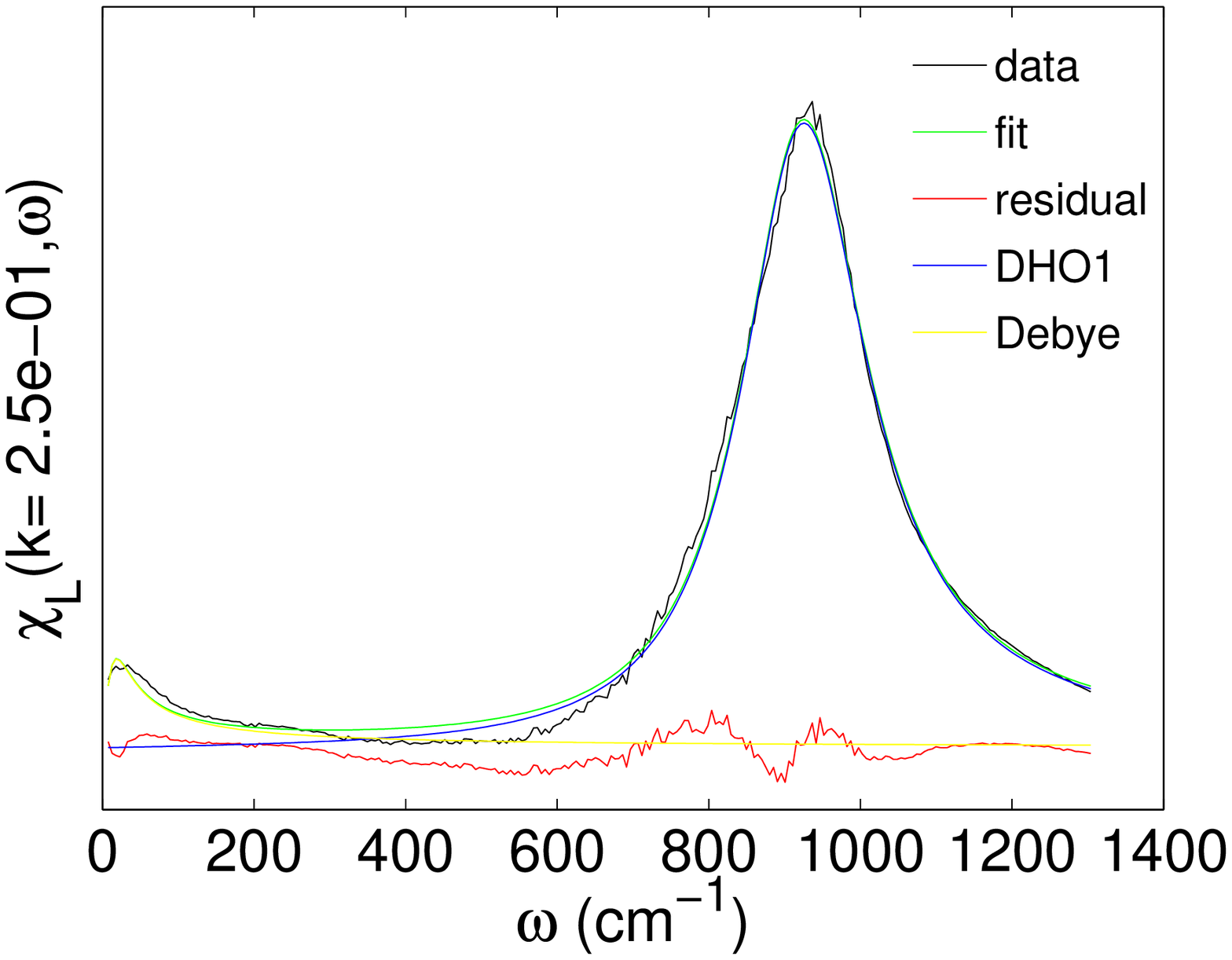}$\quad$
    \includegraphics[width=6cm]{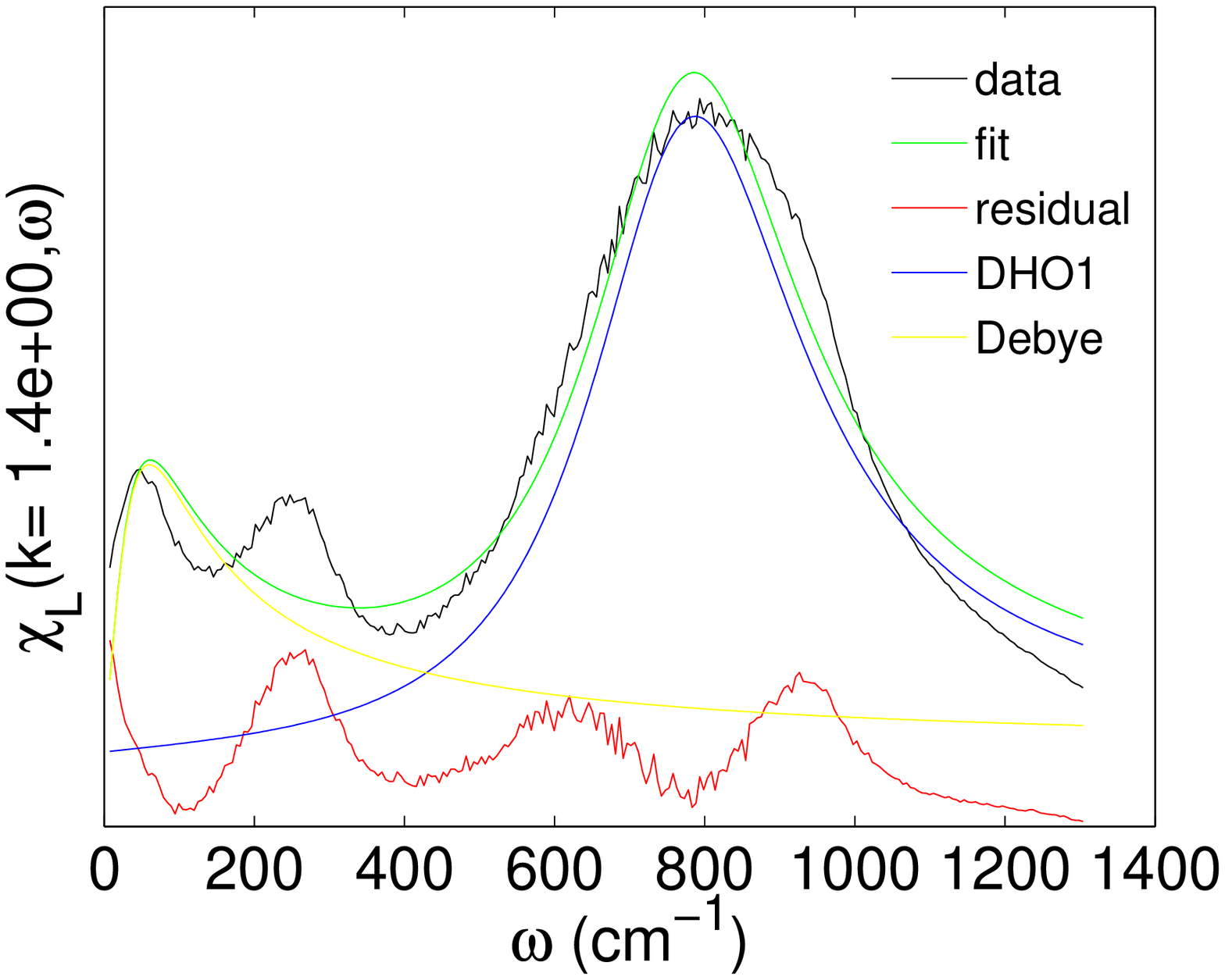}
  \caption{{\bf Example fits of the longitudinal susceptibility of TIP4P/2005f at 300K.} Fit with a Debye function and one damped harmonic oscillator at $k = .25 \Ang^{-1}$ (left) and $k = 1.4 \Ang^{-1}$ (right). Two peaks appear in the residual - the lower frequency peak is dispersive, having the same dispersion relation as the fitted peak, suggesting that it is actually part of the dispersive peak lineshape that is not captured by our lineshape function. The higher frequency peak in the residual is non-dispersive and is in the same location for both the transverse and longitudinal susceptibility.}
\end{figure}

\FloatBarrier
%------------------------------------------------------------------------------
\subsection{Supplementary info: spatial decomposition of spectra}

There are several different ways to decompose a spectra into contributions from molecules separated by distance $R$:
\begin{center}
{\bf Kirkwood dipole-sphere method}
\end{center}

This is the method we choose, which is a modification of the sphere-sphere method (see below). We start with the time-correlation function of interest : 
\begin{equation}
\phi(t) = \left\langle \sum_i \bs{\mu}_i(0) \cdot \sum_j \bs{\mu}_j(t) \right\rangle
\end{equation}
Here $\bs{\mu}$ can be replaced with any dynamical variable of interest, for instance $\bs{p}^T(\bs{k},t)$ or $\bs{j}(t)$. We omit the $k$ dependence for simplicity. 

The most straightforward way is to limit the molecules around each molecule $i$ to those in a sphere of radius $R$:
\begin{equation}
\phi(t,R) = \left\langle \sum_i \bs{\mu}_i(0) \cdot \sum_{j \in R_i} \bs{\mu}_j(t) \right\rangle
\end{equation}

This is similar to the method employed by Bopp \& Kornyshev. During the the course of a simulation molecules enter and leave each sphere, which creates noise, requiring longer averaging times. This can be improved by utilizing a smooth cutoff function:
\begin{equation}
\phi(t,R) = \left\langle \sum_i \bs{\mu}_i(0) \cdot \sum_j P_{ij}(t) \bs{\mu}_j(t) \right\rangle
\end{equation}
where 
\begin{equation}
	P_{ij} = \frac{1}{1 + e^{(R_{ij} - R)/D}}
\end{equation}
Here $D$ is a sharpness parameter determining the relative sharpness of the cutoff. We choose not to use smoothing however, finding it to be unnecessary. The result is a spectra $\chi(\bs{k},\omega, R)$ showing contributions from molecules up to radius $R$. The resulting function exhibits the expected $R \rightarrow 0$ limit, yielding only the self contribution. In the $R \rightarrow \infty$ limit, the original full response function is recovered. This function can then be numerically differentiated to show the contributions from shells of thickness $\Delta R$ centered at distance $R$.

%-------------------------------------------------
\begin{center}
{\bf Sphere-sphere method}
\end{center}
Another method discussed by Heyden, et. al. (2010) is to study the {\it autocorrelation} of the total dipole moment of a sphere of radius $R$ centered around a reference molecule, and then average over each molecule in the system.\cite{Heyden2010:12068}
\begin{equation}
\phi^P(t,R) = \sum_i \left\langle \bs{\mu}_i^P(0) \cdot \bs{\mu}_i^P(t) \right\rangle
\end{equation}
where 
\begin{equation}
	 \bs{\mu}_i^P(t) = \mathcal{N}_i(t)\sum_{j \in R_i} P_{ij}(t) \bs{\mu}_j(t) 
\end{equation}

Heyden, et. al. recommend the normalization factor $\mathcal{N}_i(t) = (1 + \sum_{j \in R_i} P_{ij}^2)^{-1/2}$ to normalize for number of molecules in each sphere. This normalization factor is chosen so that in the bulk limit ($R \rightarrow \infty$) the original full response function is obtained (in that limit $\mathcal{N}_i = 1/\sqrt{N_{\ff{mol}}}$). In the limit $R \rightarrow 0$ only the self-term contributes. Results from this method must be interpreted with a bit of care since the calculation includes all cross-correlations between molecules within the sphere centered around the reference molecule. We found that this method is more sensitive to intermolecular correlations, in particular the H-bond stretching at $\approx 250$ cm$^{-1}$ (not shown). Altogether though we found the results from this method are complementary with our results from the dipole-sphere method.
%\begin{figure}
% \includegraphics[width=8.5cm]{IR_spectra_sphere_sphere_normalized_2ns_512_TIP4P2005f.eps}
%  \caption{Distance decomposed IR spectra for TIP4P/2005f at 300 K using the technique of Heyden, et al. A smooth cut-off with a smoothing width $\sigma = .4 \Ang$ was applied. Again, the librational region is observed to have long-range contributions. }
%    \label{DistDepIR}
%\end{figure}
%We computed a distance-dependent IR spectra using the ``sphere-sphere method" (see fig \ref{DistDepIR} ) with smoothing ($D = .4 \Ang$). The IR spectra is computed using the formula: 
%\begin{equation}
%	\alpha(\omega)n(\omega) = \frac{1}{4\pi\epsilon_0}\frac{2\pi\omega^2}{3V k_B T c} \int_{-\infty}^\infty dt e^{-i\omega t} \langle \bs{M}(0)\bs{M}(t) \rangle
%\end{equation}
%The IR spectra is in qualitative agreement with the results of Heyden, et. al.\cite{Heyden2010:12068} The main difference observed when using this method is that the H-bond stretching feature at 200 cm$^{-1}$ is emphasized at $R \approx 3-6 \Ang$. 

%--------------------------------------------------------
\begin{center}
{\bf Spatial grid method}
\end{center}
To achieve higher resolution, Heyden, et. al. also introduce a spatial grid method.\cite{Heyden2010:12068} The method works by bining the molecular dipoles into grid cells. To reduce noise caused by moleules moving in and out of bins the binning is  Gaussian, meaning the dipoles are smeared with a Gaussian function. Unlike the other methods the spatial grid method does not yield the self part as $R \rightarrow 0$ so this limit requires special interpretation.

\subsection{Supplementary info: calculation of polarization vectors}
Bopp \& Kornyshev show that to get accurate results in $k$ space it is important to use the polarization vectors for each molecule rather than just the dipole moment. To calculate the polarization vector we use the method of Raineri \& Friedman.\cite{Raineri1993:8910} We utilize the defining relation for the polarization:
\begin{equation}
		\nabla\cdot\bs{P}(\bs{r},t)  = -\rho(\bs{r},t)  
\end{equation}
When transformed into Fourier space this becomes:  
\begin{equation}
		i\bs{k}\cdot\bs{P}(\bs{r},t) = -\rho(\bs{k},t) 
\end{equation}
We introduce polarization vectors for each molecule $\bs{p}_i(\bs{k})$ so that we have
\begin{equation}\label{TotalPolT}
	\bs{P}(\bs{k}) = \sum\limits_i \bs{p}_i^{N_{\ff{mol}}}(\bs{k}) e^{-i\bs{k}\cdot\bs{r_i}}
\end{equation}
where
\begin{equation}
	i\bs{k}\cdot\bs{p}_i(\bs{k}) = -\sum\limits_\alpha q_\alpha e^{-i\bs{k}\cdot\bs{r}_{\alpha i}}
\end{equation}
The molecules are indexed by $i$ and the atoms on each molecule are indexed by $\alpha$. $\bs{r}_{i\alpha} = \bs{r}_i(t) - \bs{r}_\alpha(t)$ is the distance from each atomic site to the center of mass of molecule $i$. Following Raineri \& Friedman, we use the identity 
\begin{equation}
	e^x = 1 + x\int_0^1 ds e^{xs}
\end{equation}
and taking into account the charge neutrality of each molecule we obtain
\begin{equation}
		\bs{p}_i(\bs{k}) = -\sum\limits_\alpha q_\alpha \bs{r}_{\alpha i} \int_0^1 ds e^{-i\bs{k}\cdot\bs{r}_{\alpha i} s}
\end{equation}		
\begin{equation}	\label{molPolT}	
		\bs{p}_i(\bs{k}) = \sum\limits_\alpha \frac{q_\alpha \bs{r}_{\alpha i}  }{ i \bs{k}\cdot\bs{r}_{\alpha i} } \left(e^{ i\bs{k}\cdot\bs{r}_{\alpha i} }  - 1 \right)
\end{equation} 
The transverse part is then calculated as $\bs{P}_T = \hat{\bs{k}}\times\bs{P}$, while the longitudinal component is $\bs{P}_L = \hat{\bs{k}}\cdot\bs{P}$. The longitudinal component can also be calculated more directly from: 
\begin{equation}   
	 \hat{\bs{k}}\cdot\bs{P} = \frac{i \rho(\bs{k},t)}{k} = P_L
\end{equation}
This yields the following Kubo formula for the longitudinal part of the response: 
\begin{equation}\label{chiLkt}
	\chi_L(\bs{k},\omega) = \frac{\beta}{\epsilon_0 k^2} \int_0^\infty dt \frac{d}{dt} \langle
\rho(\bs{k},t)\rho^*(\bs{k},0)\rangle e^{i\omega t}  
\end{equation}
For a system composed of point charges, the charge density is : 
\begin{equation}
	\rho(\bs{r},t) = \frac{1}{V} \sum\limits_i\sum\limits_\alpha q_{i\alpha} \delta(\bs{r} - \bs{r}_i(t) - \bs{r}_{i\alpha}(t))
\end{equation}
Again, the index $i$ runs over the molecules while $\alpha$ runs over the atomic sites on each molecule. The charge density in k-space becomes: 
\begin{equation}\label{rhokw}
	\rho(\bs{k},t) = \frac{1}{V}  \sum\limits_i\sum\limits_\alpha q_\alpha e^{-i\bs{k}\cdot(\bs{r}_i(t) + \bs{r}_{i\alpha}(t)) }  
\end{equation}
Note that this can be Taylor expanded as: 
\begin{equation}  
	\begin{aligned}
	\frac{\rho(\bs{k},t)}{k} &= \frac{1}{k} \sum\limits	_i\sum\limits_\alpha q_\alpha \sum\limits_n \frac{ (-i\bs{k} \cdot \bs{r}_{i\alpha}(t) )^n}{ n!} \\
					    &= \bs{M}(\bs{k},t) + \bs{\mathcal{Q}}(\bs{k},t) +  \bs{\mathcal{O}}(\bs{k},t) + \cdots
	 \end{aligned}
\end{equation}
Here $\bs{M}(\bs{k},t)$, $\bs{\mathcal{Q}}(\bs{k},t)$, $\bs{\mathcal{O}}(\bs{k},t)$ are contributions due to the molecular dipoles, quadrupoles and octupoles. In the limit $k \rightarrow 0$ it from supplementary equation \ref{chiLkt} it can be seen that only the dipole term contributes to the susceptibility. In the $k \rightarrow 0$ limit one obtains
\begin{equation}\label{badeqn}
  \chi_L(k,\omega) \approx \frac{\beta}{ 3 \epsilon_0 V } \int_0^\infty dt \frac{d}{dt} \langle \bs{M}_L(\bs{k},t) \cdot \bs{M}^*_L(\bs{k},0) \rangle e^{i\omega t} 
\end{equation} 
with
\begin{equation}
	\bs{M}_L(\bs{k},t) = \sum\limits_{i = 1}^{N_{\ff{mol}}} \hat{\bs{k}}\cdot\bs{\mu}_i(t) e^{i \bs{k}\cdot\bs{r}_i(t)}  
\end{equation}
This type of expression has been used previously as an approximate expression at small $k$.\cite{Bertolini:1065} However, Bopp \& Kornyshev show quite convincingly that for water the higher order multipole terms are very important, even at the smallest $k$ available in computer simulation.\cite{Kornyshev1998:1939} Neglect of the higher order terms leads to severe consequences at large $k$, and one will not recover the physical limit $\lim\limits_{k\rightarrow\infty} \chi_{L/T}(k) = 1$ unless higher order terms are included.

\subsection{Supplementary info: polarization-polarization structure factors}
\begin{figure}
 \centering   \includegraphics[width=8.5cm]{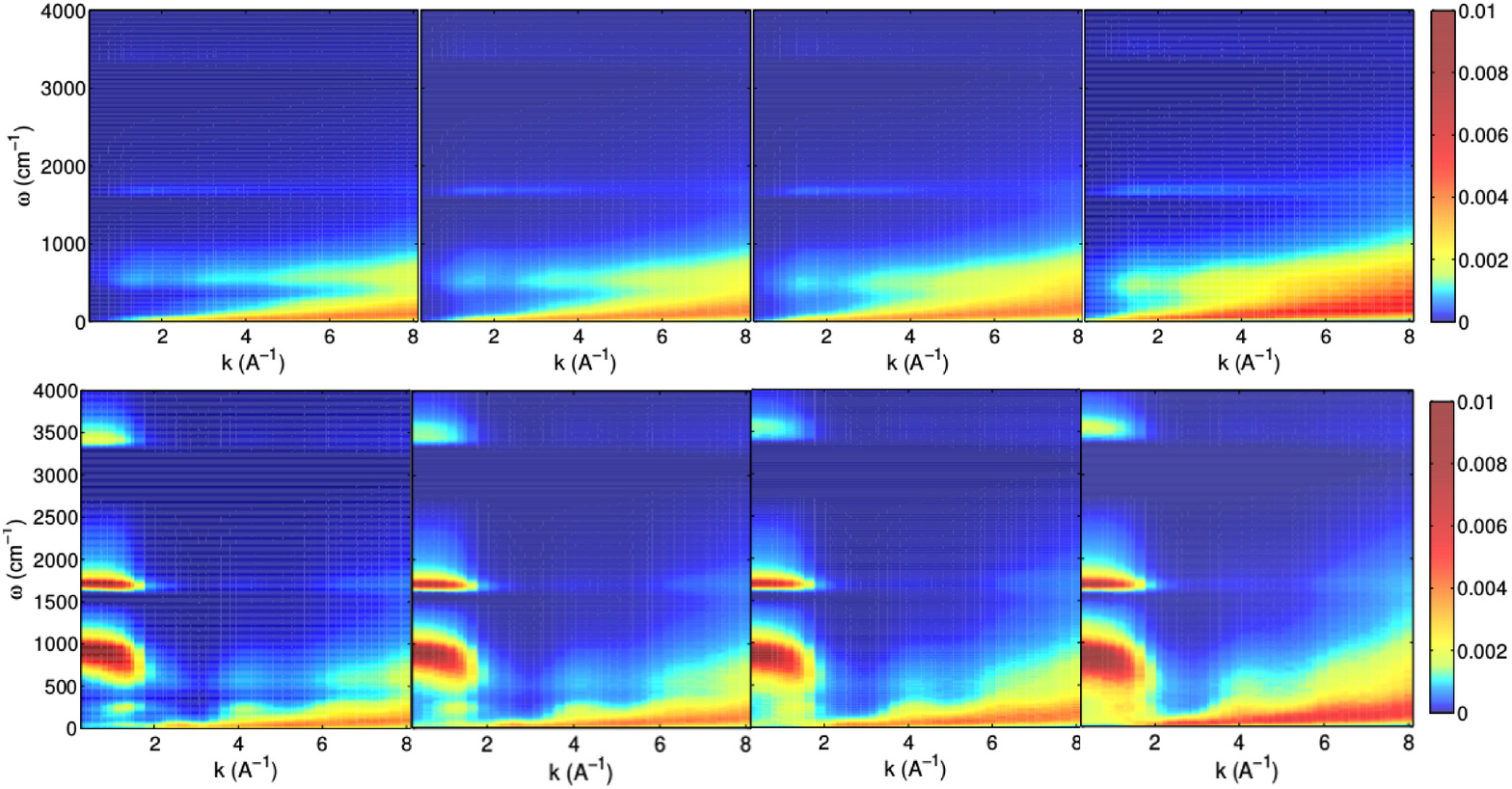}
   \caption{ Imaginary part of the longitudinal (top) and transverse (bottom) polar structure factor for TIP4P/2005f at 250 K, 300 K, 350 K, and 400 K (left to right). Note the increased intensity of the low frequency, high wavenumber intramolecular mode at higher temperatures. This is likely due to weaker H-bonding and greater freedom for inertial motion, which is responsible for this band.}
\end{figure}
To assist in visualizing $\chi_{L,T}(k,\omega)$ we introduce longitudinal and transverse ``polarization-polarization structure factors: 
\begin{equation}\label{SqqDef}
	S_{L,T}^{PP}(k,\omega) = \int_0^\infty \dot{\Phi}_{L,T}(k,t) e^{i\omega t} dt   
\end{equation}
Thus,  $\chi_{L,T}(k,\omega) = \chi_{L,T}(k,0) S_{L,T}^{PP}(k,\omega)$. These plots are shown solely because they provide a nice visual overview of the features in the nonlocal susceptibility. The main novel feature that appears in these plots is the low frequency acoustic-like mode originating at $\approx 60$ cm$^{-1}$. This mode is purely intramolecular in nature and arises from inertial rotation.\cite{Bagchi1993:133}\cite{ferraro1978} At very high wavenumbers ($k > 7)$) the relaxation is described by a rapidly decaying exponential and a Gaussian function:\cite{Bagchi1992:5126} 
\begin{equation}
	\Phi_L(k,t) = A(k) e^{t/\tau_1(k)} + B(k) e^{-(t/\tau_2(k))^2} 
\end{equation}

Gaussian relaxation yields the following equation for the imaginary part of the susceptibility: 
\begin{equation}
	\begin{aligned}
		\Phi(k,t) &= B e^{-(t/\tau)^2} \\
 \mbox{Im}\lbrace\chi(k,\omega)\rbrace &= \chi(k,0)B\frac{\sqrt{\pi}}{2}\omega\tau^2 e^{-\frac{1}{4} \tau^2 \omega^2}
	\end{aligned}	
\end{equation}
The real part is: 
\begin{equation}
 \mbox{Re}\lbrace\chi(k,\omega)\rbrace = \chi(k,0)B( \tau - \tau^2\omega F(\omega \tau / 2)) \\
\end{equation}
where $F()$ is Dawson's integral. 
The Gaussian form for the correlation function can be derived by considering a free rigid dipole subjected to Brownian kicks. In that case the relaxation function can be computed exactly.\cite{Hill1971:2322}
\begin{equation}
	\phi(t) =  \exp\left[-\frac{t}{\tau_1} + \frac{\tau_2}{\tau_1}\left\lbrace 1 - \exp\left(-\frac{t}{\tau_2}\right)\right\rbrace\right]
\end{equation}
where $\tau_1 = \xi/2k_B T$ and $\tau_2 = I/\xi$ and $\xi$ is the friction.
In either the limit $\xi \rightarrow 0$ or $t \rightarrow 0$ one obtains the Gaussian form. Thus the interpretation of the Gaussian form is that it is due to fast inertial relaxation. It has been suggested that such inertial relaxation is origin of the Poley absorption that has been found in some dipolar liquids.\cite{Hill1971:2322,H69} ``Poley absorption" appears to be used as a general term for absorption of unknown origin found in many polar liquids around 10 cm$^{-1}$ (.3 THz),\cite{H69} first described by Poley in 1955.\cite{Poley1955:337} It has been variously described as being due to fast inertial ``rattling" of molecules within their potential energy basins or as due to fast librational/inertial motion analogous to the rotational absorption of gas molecules.\cite{Bagchi1993:133,ferraro1978}

\end{document}